%% file: ms.tex
\RequirePackage{fix-cm}

\documentclass[twocolumn,epjc3, final]{svjour3}%EPJC twocolumn , draft, smallcondensed, referee

\journalname{European Physical Journal C}

\usepackage{upgreek} 
\usepackage{booktabs}
\usepackage{comment}
\usepackage{xcolor}
\usepackage{amsmath}
\usepackage{tabularx}
\usepackage{bm}
\usepackage{rotating}
\usepackage{placeins}
\usepackage{widetext}
\usepackage{times}

\newcommand{\ttwo}[1]{\mbox{T$_2$}}

\usepackage[load-configurations=abbreviations, group-digits = false]{siunitx}
\DeclareSIUnit{\TT}{T_{2}}
\DeclareSIUnit{\clight}{c}

\newcommand{\figWidth}{1}

\newcommand{\tabWidth}{0.7}

\graphicspath{{figures/}}

\begin{document}

\title{Precision measurement of the electron energy-loss function in tritium and deuterium gas for the KATRIN experiment}
\titlerunning{\parbox{9cm}{Precision measurement of the electron energy-loss function in tritium and deuterium gas for the KATRIN experiment}}
\authorrunning{KATRIN Collaboration}
\input{authors.tex}

\date{Received: date / Accepted: date}
\maketitle

\begin{abstract}
The KATRIN experiment is designed for a direct and model-independent
determination of the effective electron anti-neutrino mass via a high-precision
measurement of the tritium $\upbeta$-decay endpoint region with a sensitivity on
$m_\nu$ of \SI[per-mode=symbol]{0.2}{\electronvolt\per\clight\squared} (90\%
CL). For this purpose, the $\upbeta$-electrons from a high-luminosity
windowless gaseous tritium source traversing an electrostatic retarding spectrometer are counted to obtain an integral spectrum
around the endpoint energy of \SI{18.6}{\kilo\electronvolt}.
A dominant systematic effect of the response of the experimental setup is the energy loss of
$\upbeta$-electrons from elastic and inelastic scattering off tritium molecules
within the source.
We determined the \linebreak energy-loss function in-situ with a pulsed angular-selective
and monoenergetic photoelectron source at various tritium-source densities. The data
was recorded in integral and differential modes; the latter was achieved by
using a novel time-of-flight technique.

We developed a semi-empirical parametrization for the energy-loss function
for the scattering of \num{18.6}-\si{\keV} electrons from hydrogen isotopologs. This model was
fit to measurement data with a \SI{95}{\percent} \ttwo{} gas mixture at \SI{30}{\kelvin}, as used in the first KATRIN neutrino
mass analyses, as well as a D$_2$ gas mixture of
\SI{96}{\percent} purity used in KATRIN commissioning runs.  The achieved
precision on the energy-loss function has abated the corresponding uncertainty of $\sigma(m_\nu^2)<\SI{e-2}{\eV\squared}$ \cite{Aker2021}
in the KATRIN neutrino-mass measurement to a subdominant level.

\end{abstract}


\tableofcontents

\input{introduction}
\input{energyloss}
\input{measurementPrinciple}
	\input{integralMode}
	\input{differentialMode}
	\input{pileup}
	\input{background}
\input{fitResults}
\input{errorPropagation}
\input{deuteriumResults}
\input{summary}

\section*{Acknowledgments} 
We acknowledge the support of Helmholtz Association, \linebreak Ministry for Education and Research BMBF (5A17PDA, \linebreak 05A17PM3, 05A17PX3, 05A17VK2, and 05A17WO3),\linebreak Helmholtz Alliance for Astroparticle Physics (HAP),  Helmholtz Young Investigator Group (VH-NG-1055), and Deutsche Forschungsgemeinschaft DFG (Research Training \linebreak Groups GRK 1694 and GRK 2149, and Graduate School GSC 1085 - KSETA) in Germany; Ministry of Education, Youth and Sport (CANAM-LM2015056, LTT19005) in the Czech Republic; Ministry of Science and Higher Education of the Russian Federation under contract 075-15-2020-778; and the United States Department of Energy through grants DE-FG02-97ER41020, DE-FG02-94ER40818, DE-SC0004036, DE-FG02-97ER41033, DE-FG02-97ER41041,    DE-SC0011091 and DE-SC0019304, Federal Prime Agreement DE-AC02-05CH11231, and the National Energy Research Scientific Computing Center.
\appendix
\addcontentsline{toc}{section}{Appendix}
\input{appendix}
\bibliography{bibliography}
\end{document}

%% file: authors.tex
% autogenerated by authorlist.py from input file 'authorsList_by_Alan_copy.tex' using svjour3 format

% Affiliations:
\institute{%
Tritium Laboratory Karlsruhe~(TLK), Karlsruhe Institute of Technology~(KIT), Hermann-von-Helmholtz-Platz 1, 76344 Eggenstein-Leopoldshafen, Germany\label{a}
\and Institute for Data Processing and Electronics~(IPE), Karlsruhe Institute of Technology~(KIT), Hermann-von-Helmholtz-Platz 1, 76344 Eggenstein-Leopoldshafen, Germany\label{b}
\and Institute for Astroparticle Physics~(IAP), Karlsruhe Institute of Technology~(KIT), Hermann-von-Helmholtz-Platz 1, 76344 Eggenstein-Leopoldshafen, Germany\label{c}
\and Institute of Experimental Particle Physics~(ETP), Karlsruhe Institute of Technology~(KIT), Wolfgang-Gaede-Str. 1, 76131 Karlsruhe, Germany\label{d}
\and Institute for Nuclear Research of Russian Academy of Sciences, 60th October Anniversary Prospect 7a, 117312 Moscow, Russia\label{e}
\and Institut f\"{u}r Kernphysik, Westf\"{a}lische Wilhelms-Universit\"{a}t M\"{u}nster, Wilhelm-Klemm-Str. 9, 48149 M\"{u}nster, Germany\label{f}
\and Technische Universit\"{a}t M\"{u}nchen, James-Franck-Str. 1, 85748 Garching, Germany\label{g}
\and Max-Planck-Institut f\"{u}r Physik, F\"{o}hringer Ring 6, 80805 M\"{u}nchen, Germany\label{h}
\and Department of Physics and Astronomy, University of North Carolina, Chapel Hill, NC 27599, USA\label{i}
\and Triangle Universities Nuclear Laboratory, Durham, NC 27708, USA\label{j}
\and Institute for Nuclear and Particle Astrophysics and Nuclear Science Division, Lawrence Berkeley National Laboratory, Berkeley, CA 94720, USA\label{k}
\and Department of Physics, Faculty of Mathematics and Natural Sciences, University of Wuppertal, Gau\ss{}str. 20, 42119 Wuppertal, Germany\label{l}
\and Departamento de Qu\'{i}mica F\'{i}sica Aplicada, Universidad Autonoma de Madrid, Campus de Cantoblanco, 28049 Madrid, Spain\label{m}
\and Center for Experimental Nuclear Physics and Astrophysics, and Dept.~of Physics, University of Washington, Seattle, WA 98195, USA\label{n}
\and Nuclear Physics Institute of the CAS, v. v. i., CZ-250 68 \v{R}e\v{z}, Czech Republic\label{o}
\and Laboratory for Nuclear Science, Massachusetts Institute of Technology, 77 Massachusetts Ave, Cambridge, MA 02139, USA\label{p}
\and Department of Physics, Carnegie Mellon University, Pittsburgh, PA 15213, USA\label{q}
\and IRFU (DPhP \& APC), CEA, Universit\'{e} Paris-Saclay, 91191 Gif-sur-Yvette, France \label{r}
\and Max-Planck-Institut f\"{u}r Kernphysik, Saupfercheckweg 1, 69117 Heidelberg, Germany\label{s}
\and Institut f\"{u}r Physik, Humboldt-Universit\"{a}t zu Berlin, Newtonstr. 15, 12489 Berlin, Germany\label{t}
}
\thankstext{email1}{mail: rodenbeck@wwu.de}
\thankstext{email2}{mail: lschimpf@wwu.de}
% Authors:
\author{%
M.~Aker\thanksref{a}
\and A.~Beglarian\thanksref{b}
\and J.~Behrens\thanksref{c,d}
\and A.~Berlev\thanksref{e}
\and U.~Besserer\thanksref{a}
\and B.~Bieringer\thanksref{f}
\and F.~Block\thanksref{d}
\and B.~Bornschein\thanksref{a}
\and L.~Bornschein\thanksref{c}
\and M.~B\"{o}ttcher\thanksref{f}
% ...
}

%% file: introduction.tex
\section{Introduction}
\begin{figure*}[tbp]
	\centering
	\includegraphics[width=0.95\textwidth]{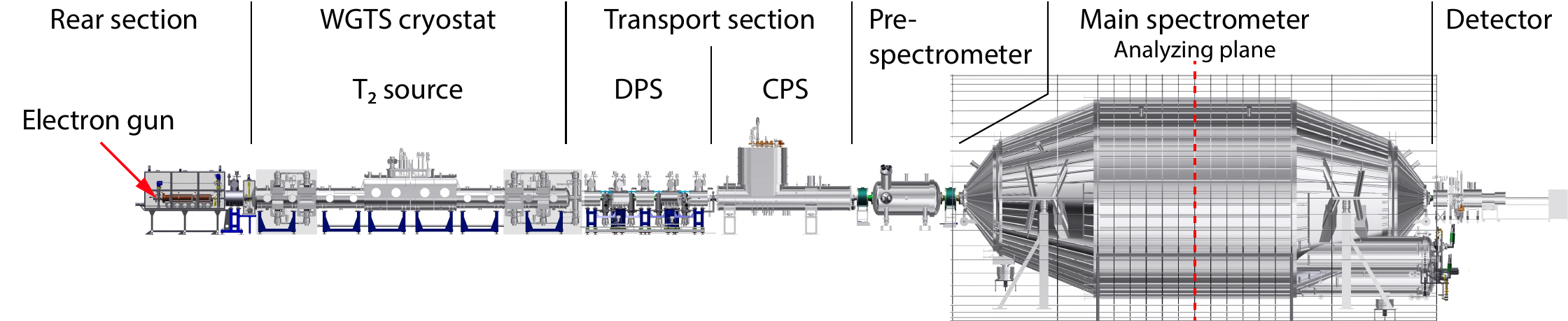}
	\caption{Overview of the KATRIN experiment. The main components are from left to right: The rear section containing calibration and monitoring systems as well as the electron gun (see Fig.~\ref{fig:egun_drawing}) used in this work; the \num{10}-\si{\m}-long windowless gaseous tritium source (WGTS) with differential pumps on both sides; the transport section consisting of a differential (DPS) and cryogenic pumping section (CPS); the spectrometer and detector section with the pre- and main spectrometer, and the silicon detector. The overall length of the experimental setup is more than \SI{70}{\meter}.
		\label{fig:katrin}}
\end{figure*}

The KArlsruhe TRItium Neutrino (KATRIN) experiment\linebreak aims to determine the
effective electron anti-neutrino mass in a model-independent way by examining
the kinematics of tritium $\upbeta$-decays. The observable $m^2_\nu = \sum_{i}{ \left | U_{\mathrm{e}i}\right |^2 m^2_{i}}$ is the squared incoherent sum of neutrino-mass eigenstates $m_{i}$\linebreak weighted by their contribution $U_{\mathrm{e}i}$ to the electron \linebreak anti-neutrino.
The target sensitivity for the neutrino-mass measurement in KATRIN is
0.2~eV/c$^2$ (at 90\% CL) with three live-years of data~\cite{KAT04}. The
$5\sigma$ discovery potential is 0.35~eV/c$^2$. This requires a precise control
of all systematic effects.
The experiment is designed for a high-precision spectral shape measurement of
\ttwo{} $\upbeta$-decay electrons around the endpoint of 18.6~keV.  An overview
of the KATRIN experiment is shown in Fig.~\ref{fig:katrin}. The setup \cite{Aker2021design} includes a
high-activity Windowless Gaseous Tritium Source (WGTS) and a high-resolution
electrostatic retarding spectrometer of the MAC-E (Magnetic Adiabatic Collimation with an Electrostatic filter) type \cite{Beamson1980,LOBASHEV1985,Pic92}.
Molecular tritium gas at 30~K is continuously injected through the capillaries
at the center of the WGTS and pumped out at both ends. This allows a nominal
steady-state column density (i.e. the integrated nominal source density $\rho_0(z)$ along the length $d$ of the source cryostat) $\rho_0d=\SI{5e17}{\per\centi\meter\squared}$ resulting in an activity of \SI{1.7e11}{\becquerel}
with a stability better than \SI{ 0.1}{\percent\per\hour} \cite{Aker2021design}.

In order to prevent tritium from entering the spectrometer section which would
induce background in the measurement, the transport section reduces the tritium
flow by at least 14 orders of magnitude \cite{Friedel2019}. This is
achieved with a differential pumping section \cite{Aker2021design,Marsteller2021}, which comprises turbo-molecular
pumps followed by a cryogenic pumping section that makes use of an argon frost
layer to adsorb tritium cryogenically \cite{Aker2021design,Roettele2017}. The spectrometer section consists of the
pre- and the main spectrometer. The pre-spectrometer rejects\linebreak low-energy
electrons, which reduces the electron flux into the main spectrometer. The final
precision discrimination of the electron energy is performed in the analyzing
plane at the center of the main spectrometer with a resolution of
\SI{2.77}{\eV} \cite{Aker2021design} for \num{18.6}-\si{\kilo\electronvolt} electrons with isotropic angular distribution.

The pre- and main spectrometer are MAC-E type high-pass filters,
which can only be traversed by electrons with longitudinal kinetic energy
higher than the preset potential. The isotropically emitted $\upbeta$-electrons
are adiabatically collimated to a longitudinal motion inside the
spectrometer. This is achieved by a gradual decrease of the magnetic field
strength $B$ from the entrance of the spectrometer towards its center, conserving
the magnitude of the $\upbeta$-electron's magnetic moment in the cyclotron motion
$\mu =E_\perp/B$ \cite{Beamson1980}, with $E_\perp$ being the transverse component of the electron's kinetic energy with respect to the magnetic field lines.  Varying the electric potential of the spectrometer allows the
energy region around the endpoint of the tritium $\upbeta$-decay to be scanned
as an integral spectrum, i.e. the rate of electrons with kinetic energy above the
set filter potential \cite{Aker2021}.

Electrons passing the main spectrometer are re-ac\-cel\-er\-a\-ted by the main spectrometer potential and a post-ac\-cel\-er\-a\-tion of \SI{10}{\kV} at the focal-plane detector (FPD) system and are then
counted by a 148-pixel silicon PIN detector~\cite{Ams15} shown at the far right
in Fig.~\ref{fig:katrin}.  An \num{18}-\si{\keV}-wide selection window (\SI{14}{\keV} to \SI{32}{\keV}) around
the \num{28}-\si{\keV} electron energy peak is chosen to minimize systematic effects in
counting efficiencies \cite{Aker2021}.

The observable $m_\nu^2$ is determined by fitting the recorded integral spectrum
with a model that comprises four parameters: the normalization, the
endpoint energy, the background rate, and $m_\nu^2$ \cite{Kleesiek2019}.  The
model is constructed from the shape of the $\upbeta$-decay spectrum and
the response of the experimental setup. The main components of the response are
the transmission function of the main spectrometer and the energy loss of
electrons from elastic and inelastic scatterings in the \ttwo{} source. The
latter is the focus of this work.

At the nominal source density, approximately \SI{60}{\percent} of all
electrons scatter inelastically and lose energies between
$\approx\,$\SI{11}{\electronvolt} and \SI{9.3}{\kilo\electronvolt}. The
upper limit of this energy transfer arises due to the fact that the primary and secondary electrons from the ionization process
are indistinguishable in the measurement and always the higher energetic electron is measured.
Minuscule energy losses can result in electrons with energies close to the
endpoint downgraded to lower energies in the spectrum fit window. Therefore,
the energy-loss function needs to be known with high precision in order to meet the systematic uncertainty budget of\linebreak $\sigma(m_\nu^2)<\SI{7.5e-3}{\eV\squared}$ \cite{KAT04} reserved for this individual systematic.

Theoretical differential cross sections for \num{18.6}-\si{\keV} electrons
scattering off molecular tritium are not available at the required precision for
the $m_\nu^2$ measurements.  While data from energy-loss measurements for
gaseous tritium or deuterium from the former neutrino mass
experiments in Troitsk and Mainz~\cite{Ase00,Abd17} exist, the precision is not
sufficient to achieve the KATRIN design sensitivities. Other more precise
experimental data on the energy losses of electrons with energies near the
tritium $\upbeta$-decay endpoint energy are only available for molecular
hydrogen as the target gas~\cite{Abd17,Gei64,Uls72}. In this paper we report the
results of the in-situ measurements of the energy-loss function in the KATRIN
experiment.

We used a monoenergetic and angular-selective electron gun, of the type described
in \cite{Beh16}, mounted in the rear section (far left in Fig.~\ref{fig:katrin}),
which allowed us to probe the response of the entire KATRIN setup, including the energy loss in tritium gas.

We begin this paper in Sec.~\ref{sec:eloss} with a brief introduction to existing energy-loss function models and continue with the description of the novel semi-empirical parametrization developed in this work. In Sec.~\ref{sec:measurements}, the measurement approaches of the integral as well as the novel differential time-of-flight measurements are explained, including a description of the working principle of the electron gun used for these measurements. The analysis of the tritium data using a combined fit is presented in Sec.~\ref{sec:Analysis} including a detailed discussion of the systematic uncertainties of the measurements.
Additional measurement results for the energy-loss function in deuterium gas
are provided in Sec.~\ref{sec:D2}.
We conclude this paper in Sec.~\ref{sec:summary} by summarizing and discussing our results in the context of the neutrino-mass-sensitivity goal of KATRIN.

%% file: energyloss.tex
\section{Energy-loss function}
\label{sec:eloss}
Multiple processes contribute to the energy loss of electrons traversing
molecular tritium gas. The median energy loss from elastic scattering amounts to $\overline{\Delta
  E}_{\mathrm{el}}=\SI{2.3}{\milli\electronvolt}$ \cite{Kleesiek2019}, which is
negligible in the KATRIN measurement. The predominant processes for the KATRIN
experiment are inelastic scatterings, resulting in electronic excitations in combination with rotational and
vibrational excitations of the molecule, ionization, and molecular dissociation.

Data from detailed measurements is only available for the scattering of
\num{25}-\si{\kilo\electronvolt} electrons on molecular hydrogen
gas~\cite{Gei64,Uls72}; these direct measurements of the energy-loss function
were made with energy resolutions down to 40~meV. 
In these measurements, the contribution of three different groups of lines can be discerned, which are created from the excitations of the
$(\mathrm{2p}\sigma\ ^1\Sigma^+_\mathrm{u})$, $(\mathrm{2p}\pi\ ^1\Pi_\mathrm{u})$, and $(\mathrm{3p}\pi\ ^1\Pi_\mathrm{u})$ molecular states around \SI{12.6}{\electronvolt} and \SI{15}{\electronvolt}, respectively.

Aseev et al.~\cite{Ase00} and Abdurashitov et al. \cite{Abd17} report on the measurements of energy losses of
electrons in gaseous molecular hydrogen, deuterium, and tritium. The shape of the energy-loss function was evaluated by fitting an
empirical model to the integral energy spectra obtained with a mono-energetic electron source  which generated a beam of electrons with kinetic energies near the endpoint energy of the tritium $\upbeta$-decay. Because of the
low energy resolution of several eV, the shape of the energy-loss function was
coar\-se\-ly approximated by a Gaussian to represent electronic excitations and
dissociation, and a one-sided Lorentzian to represent the continuum caused
by ionization of the molecules \cite{Ase00}.

\subsection{New Parametrization}
\label{sec:parametrization}
The high-quality data from the first KATRIN energy-loss measurements described in
Sec.~\ref{sec:measurements} allows us to improve the parametrization used in
Aseev et al.~\cite{Ase00} and Abdurashitov et al.~\cite{Abd17}.  While the
experimental energy resolution is not sufficient to resolve individual molecular
states, the combined contribution of each of the three groups of states can
clearly be discerned in the KATRIN data.

A new parametrization of the energy-loss function was developed to describe the
inelastic scattering region between about 11~eV and 15~eV using three Gaussians,
each of which is approximating one group of molecular states. The ionization
continuum beyond this energy region is described by the relativistic
binary-encounter-dipole (BED) model developed by Kim et
al.~\cite{Kim2000}. While the parameters required by this model are only
available for the ionization of H$_2$-molecules \cite{Kim94}, by taking into
account the ionization thresholds for the different isotopologs \cite{Wec99}
\begin{equation}
\begin{split}
E_\mathrm{i}(\mathrm{H}_2)&=\SI{15.433}{eV}\\
E_\mathrm{i}(\mathrm{D}_2)&=\SI{15.470}{eV}\\ E_\mathrm{i}(\mathrm{T}_2)&=\SI{15.486}{eV}\,,
\end{split}
\end{equation}
the shape of the BED model is a good
representation for the tritium data, as can be seen from the fit result in Sec.~\ref{sec:combinedFitmodel}. The
new parametrization of the full energy-loss function is written as:
\begin{equation}
\label{eq:katrinFitModel}
f(\Delta E) = \begin{cases}
    \sum_{j=1}^3 a_j \exp\left( -\frac{(\Delta E-m_j)^2}{2\sigma_j^2} \right)  
            &: \Delta E \le E_\mathrm{i}\\
          \frac{f(E_\mathrm{i})}{f_{\rm BED}(E_\mathrm{i}) } \cdot f_{\rm BED}(\Delta E)  &: \Delta E > E_\mathrm{i},
\end{cases}
\end{equation}
where $\Delta E$ is the energy loss and $a_j$, $m_j$, and $\sigma_j$ are the amplitude, the mean, and the width of the three Gaussians, respectively. $f_{\rm BED}(\Delta E)$ is the functional form of the BED model as given in~\cite{Kim2000} and $E_\mathrm{i}$ is the junction point between the two regions given by the ionization threshold. For a \linebreak smooth continuation of the model at the junction, the BED function $f_{\rm BED}(\Delta E)$ is normalized to the local value $f(E_\mathrm{i})$ of the Gaussian components at that position. 

%% file: measurementPrinciple.tex
\section{Measurements}
\label{sec:measurements}
The energy-loss function $f(\Delta E)$ (Eq.~\ref{eq:katrinFitModel}) describes the electron energy losses $\Delta E$ from scattering inside the source, which distort the shape of the response function.
By
measuring the response function, it is possible to determine $f(\Delta E)$. For
this, a quasi-monoenergetic and angular-selective photoelectron source (``electron gun"), located at the end of the rear section (see
Fig.~\ref{fig:katrin}), is used.
Guiding the quasi-mo\-no\-e\-ner\-ge\-tic beam --- at a
pitch angle of approx. $\theta=\SI{0}{\degree}$ between the magnetic field lines
and the electrons' momentum vector --- through the WGTS allows the investigation
of the energy loss from scatterings with the source gas molecules stabilized at
\SI{30}{\kelvin}. Measuring the electron rate at the focal-plane detector as a
function of the electron surplus energy $E_\mathrm{s}$ at the analyzing plane (see Eq.~
\ref{eq:surplusEnergy}) yields the response function of the setup.

The working principle of the electron gun and a general description of the measurement strategy are provided in the following. This is followed by a discussion of the measurement data taken in the two different measurement modes (integral and differential) as well as two important systematic effects in the measurements (pile-up and background).

\begin{figure}[tbp]
	\centering
	\includegraphics[width=\figWidth\linewidth]{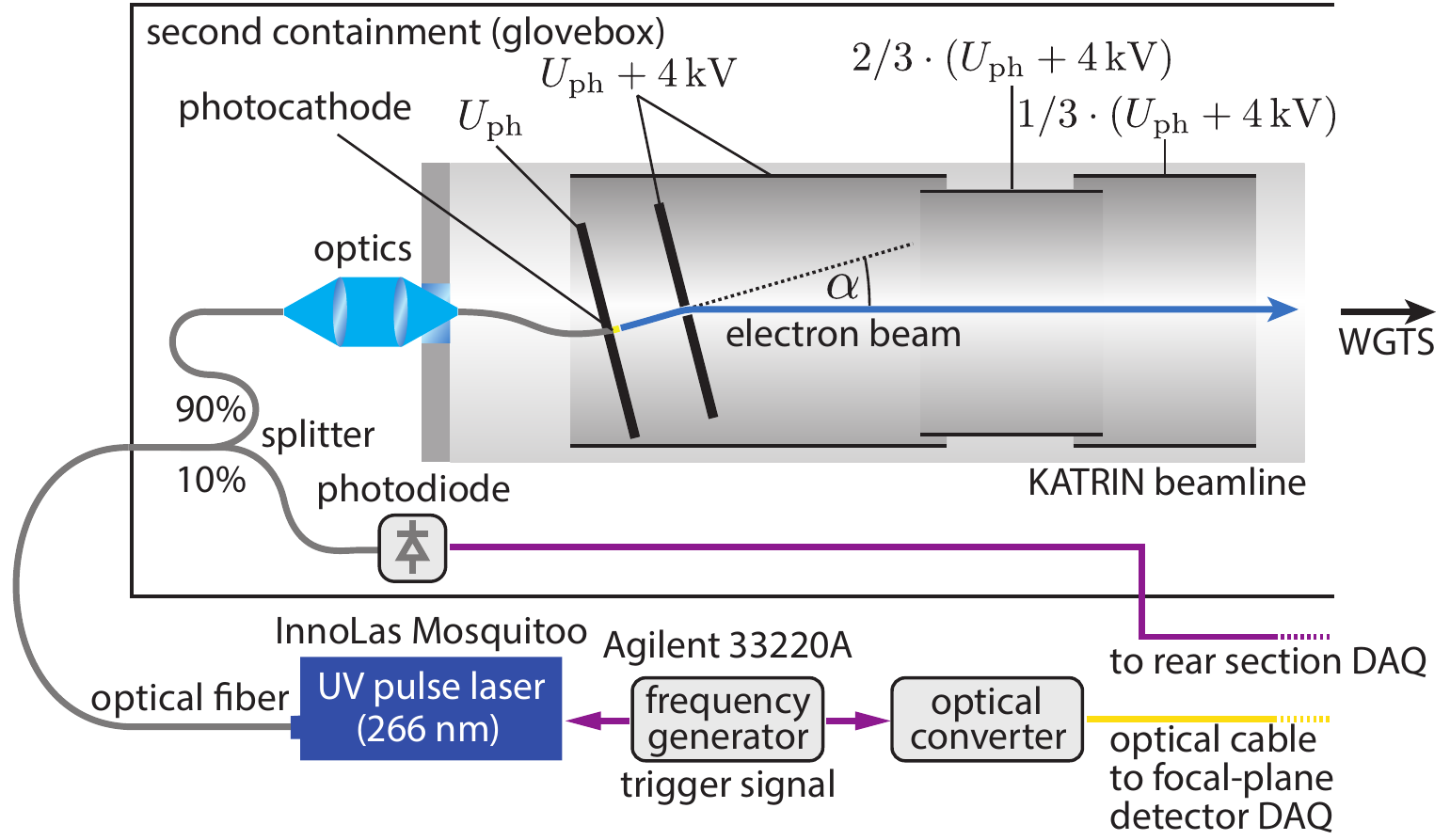}
	\caption{A simplified schematic drawing of the electron gun, including the acceleration electrodes as well as the optical setup used to generate the photoelectrons.}
	\label{fig:egun_drawing}
\end{figure}

\paragraph{Electron gun}
A schematic drawing of the electron gun is provided in
Fig.~\ref{fig:egun_drawing}. The electrons are generated by photoelectric
emission when ultraviolet light is shone through an approximately
\num{30}-\si{\nm}-thick gold photocathode, which is installed inside two electrically charged parallel plates. The photoelectrons are accelerated by a
potential difference of \SI{4}{\kV} between the plates separated by
\SI{10}{\milli\meter}; the electrons exit the setup through a hole in the front
plate (see Fig.~\ref{fig:egun_drawing}). This first non-adiabatic acceleration
collimates the beam of photoelectrons in a cosine distribution \cite{Pei_2002} initially. By tilting the plates by the angle $\alpha$, well-defined pitch
angles $\theta$ can be obtained. A pitch angle of $\theta=\SI{0}{\degree}$, which is reached by aligning the plates with the magnetic field lines, is used in the measurements. The generated electrons are further accelerated by a
cascade of cylinder electrodes to the desired kinetic energy. The working
principle is explained in more detail in \cite{Beh16}. The energy profile of the
generated beam depends on the work function $\Phi$ of the photocathode and the
wavelength $\lambda$ of the light source.

For the measurements in this work, a \SI{266}{\nano \meter} pulsed UV
laser\footnote{InnoLas Mosquitoo Nd:YVO$_4$ \SI{1064}{\nm} (frequency
  quadrupled).} with pulse widths of less than \SI{18}{\nano \second} (FWHM) is
used. The Q-switch of the laser can be externally triggered, which allows the
synchronization of the creation time of the electron pulses with the detector
system. This allows the time-of-flight (TOF) of the
signal electrons to be measured. The TOF is used for a differential analysis of
the data (see Sec.~\ref{sec:differential}).

The photon energy of the monochromatic laser light\linebreak (${h\nu=\SI{4.66}{\eV}}$) is
only \SI{0.22}{\eV} above the work function $\Phi=\SI{4.44}{\electronvolt}$
\cite{PhDSack2020} of the gold photocathode, which results in a measured
energy spread of $\sigma_\mathrm{E}<\SI{90}{\meV}$.

To generate electrons with well defined kinetic energies close to the tritium
endpoint, voltages down to \SI{-21}{\kilo\volt} can be applied to the
photocathode and cylinder electrodes. The photocathode potential
$U_{\mathrm{ph}}$ is varied to produce electrons with different surplus energies
$E_\mathrm{s}$ with respect to the negative main spectrometer retarding
potential $U_0$:
\begin{align}
\label{eq:surplusEnergy}
E_\mathrm{s} & = q  \cdot U_{\mathrm{s}}+h \nu-\Phi_i = q \cdot \left (U_{\mathrm{ph}} - U_0 \right )+h \nu-\Phi_i\,,
\end{align}
taking into account the additional initial energy of the electrons given by the difference of the photon energy $h\nu$ and the work function $\Phi_i$ of the electrons populating different energy levels in the solid (neglecting further solid-state effects).
The total initial kinetic energy of the electrons is given as $E_{\mathrm{kin}}=q\cdot U_{\mathrm{ph}}$.

\paragraph{Measurement approach}
To resolve the fine structures of the response function, small voltage steps on the
order of \SI{0.1}{\eV} are required over the analysis interval of
$E_\mathrm{s}=$\SIrange{-5}{60}{\electronvolt}. Multiple fast voltage sweeps in
alternating directions are preferred to compensate for systematic uncertainties
associated with the scan direction and long-term instabilities of the setup.
A single high-voltage setpoint adjustment of the main spectrometer requires more
than \SI{10}{\s} to stabilize, which does not allow repeated measurements within
a reasonable time.  For faster measurements, the surplus energy of the electron
beam is modified by performing voltage sweeps of $U_{\mathrm{ph}}$ while the
filter potential of the main spectrometer is kept fixed at
$U_0=-\SI{18575}{\volt}$. The electron energy is chosen to be slightly above the
tritium endpoint energy to avoid $\upbeta$-electron backgrounds but close to the
region of interest to minimize effects from the energy dependence of the
scattering cross section.

Changing the kinetic energy of the electrons results in a small change of the total
inelastic scattering cross section $\sigma^\mathrm{tot}_\mathrm{inel}$ of up to \SI{0.27}{\percent} over
the scanned energy range. This is considered later in the data analysis.

Each sweep (called ``scan'' in the following) took \SI{30}{\minute} and was
repeated in alternating scanning directions for approximately \SI{12}{\hour}.
The obtained rates as a function of the continuous voltage ramp were binned to
obtain discrete energy values for the analysis. The data taking was performed in
integral and differential modes, which are described in more detail in the
following.

%% file: integralMode.tex
\subsection{Integral measurements}
In the standard KATRIN measurement mode, only electrons with high enough surplus
energies to overcome the main spectrometer retarding potential reach the
detector. By \linebreak changing the kinetic energy of the electrons and keeping the
retarding potential at a fixed value, the integral response function was measured.
A set of integral measurements at three different non-zero column densities as
well as one reference measurement at zero column density (see
Tab.~\ref{tab:integralMeasurements}) were performed. The pulse frequency of the
laser was set to \SI{100}{\kilo\hertz}, which results in an estimated mean value
of 0.05 generated electrons per light pulse.

\begin{table}[tbp]
	\centering
	\caption{A summary of the number of scans $\Sigma$ performed at different
      column densities relative to the nominal value $\rho_0d$. The
      corresponding scattering probability $\mu$ is also shown. The average
      number of counts $<N_0>$ per \num{50}-\si{\milli\volt} bin for the unscattered
      electrons at $E_s\in\left[\SI{2}{\eV},\SI{10}{\eV}\right]$ is provided for
      the integral dataset, as well as the sum of all unscattered electrons
      $N_0$ at $E_s\in\left[\SI{-1}{\eV},\SI{1}{\eV}\right]$ for the differential
      dataset.}
	\begin{tabular}{c c c c}
		\multicolumn{4}{c}{Integral}\\
\toprule
		Column density / $\rho_0d$ &$\mu$&$\Sigma$ & $<N_0>$\\
\midrule
		\SI{0}{\percent} & 0.00 & 28 & 204806\\
		\SI{14}{\percent} & 0.25 &14 & 88002\\
		\SI{41}{\percent} & 0.75 &26 & 112655\\
		\SI{86}{\percent} & 1.56 & 31 & 62191\\
\bottomrule
 & & &  \\
		\multicolumn{4}{c}{Differential}\\
\toprule
		Column density / $\rho_0d$ &$\mu$&$\Sigma$ & $N_0$\\
\midrule
\SI{15}{\percent} & 0.27 &33 & 565316 \\
\SI{22}{\percent} & 0.41 &23 & 380633\\
\SI{39}{\percent} & 0.72 &23 & 267829\\
\SI{84}{\percent} &1.52  & 28 & 154460\\
\bottomrule
	\end{tabular}
	\label{tab:integralMeasurements}
\end{table}

\begin{figure}[tbp]
	\centering
	\includegraphics[width=\figWidth\linewidth]{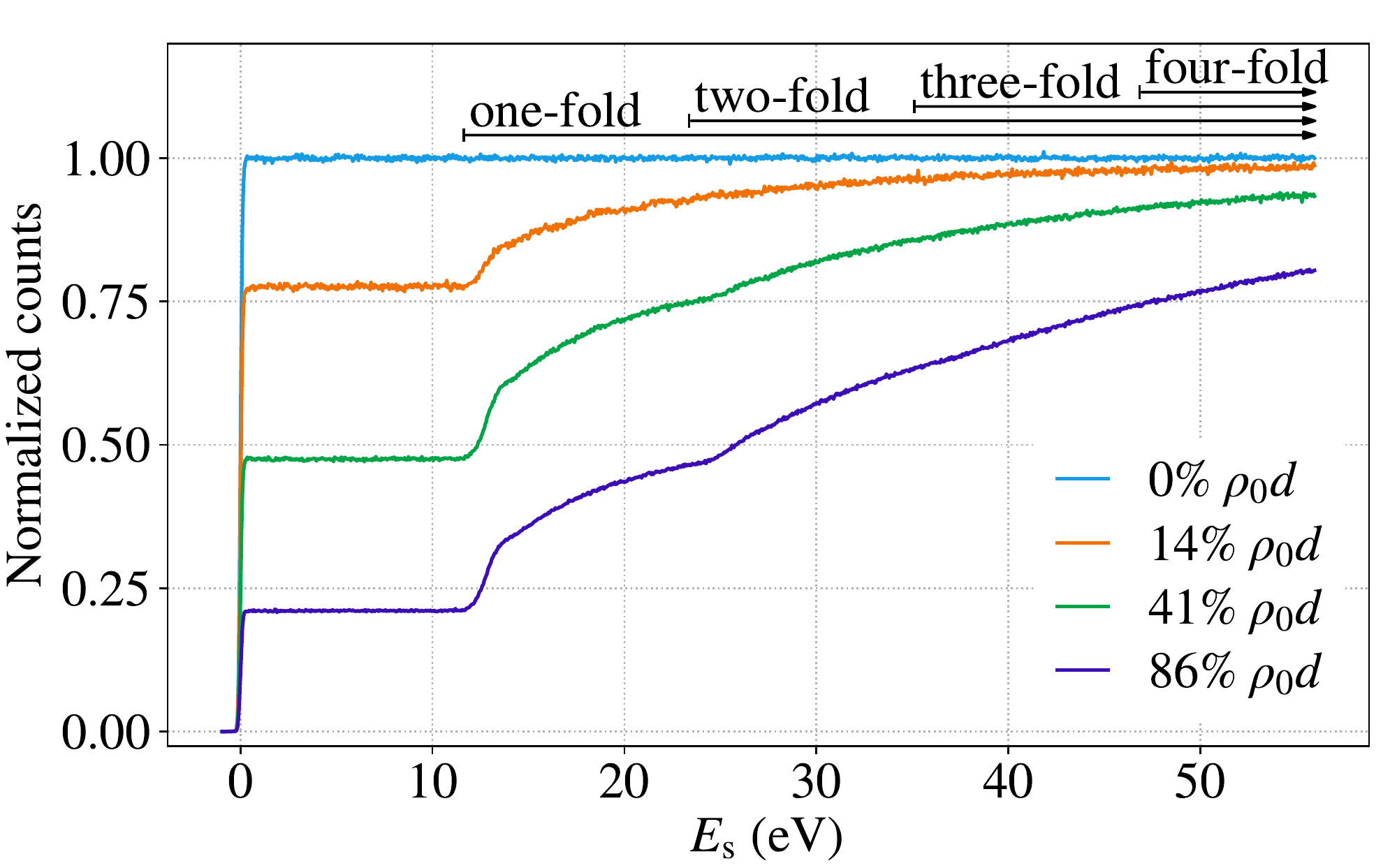}
	\caption{The measured response functions in integral mode at different fractions of the nominal
      column density $\rho_0 d$. The response functions are normalized by the
      electron rate of the reference measurement with an empty source (blue
      curve). The arrows indicate the energy region where $n$-fold scattering
      takes place.}
	\label{fig:dataOnly_integral}
\end{figure}

The individual scans were both corrected for rate intensity fluctuations and
detector pile-up (see Sec.~\ref{sec:pileup}).  The former are caused by
fluctuations of the laser intensity, which is stable to
\SI{1.2}{\percent\per\hour}. The light intensity is continuously monitored by a
photodiode connected to a fiber splitter, which is installed just before the light is coupled into the vacuum system of the electron gun (see Fig.~\ref{fig:egun_drawing}). The light intensity
correction is done by dividing the measured FPD rate by the relative deviation
of the light intensity to its mean intensity. The precision of the measured light intensity with this monitoring system is \SI{0.4}{\percent} and is propagated into the uncertainties of the correction.
Data from scans at the same column density are accumulated (Fig. ~\ref{fig:dataOnly_integral}).
The resulting integral response functions are superpositions of $n$-fold scattering functions, as indicated in the figure by arrows above the measurement data.

\begin{comment}
But since multiple scatterings can take place on the measurement range of up to \SI{60}{\electronvolt}, the measured response function is always an superposition of n-fold convolutions of the energy-loss function.
\end{comment}

%% file: differentialMode.tex
\subsection{Differential (time-of-flight) measurements}
\label{sec:differential}

The time of each trigger pulse for the laser is saved in the detector data
stream and used to define the electron-emission time at the electron gun. For
each event at the detector, its time difference to the laser pulse is
calculated. The time difference corresponds to the time-of-flight (TOF) of the
electron through the KATRIN beamline from the electron gun to the detector,
including delays for the signal propagation and processing on the order of
\SI{1}{\micro\second}. The knowledge of the electron's time-of-flight can be used
as additional information on its kinetic energy.

The negative retarding potential in the main spectrometer $U_0$ acts as a
barrier for the electrons, slowing them down and only allowing electrons with
surplus energies $E_\mathrm{s}> 0$ to pass through (high-pass filter). The
higher the electrons' surplus energy, the less they are slowed down inside the
main spectrometer; connecting their flight time through the main
spectrometer $\tau$ to their surplus energy by $\tau \sim
\frac{1}{\sqrt{E_\mathrm{s}}}$ \cite{Steinbrink2013}.

Selecting only electrons with $\tau > \tau_{\mathrm{cut}}$ is equivalent to a
low-pass filter on $E_{\mathrm{s}}$ \cite{Bonn1999}. Applying this TOF
selection, the high-pass filter main spectrometer is transformed into a narrow
band-pass filter for measuring the differential energy spectrum.

\begin{figure}[tbp]
	\centering
	\includegraphics[width=\figWidth\linewidth]{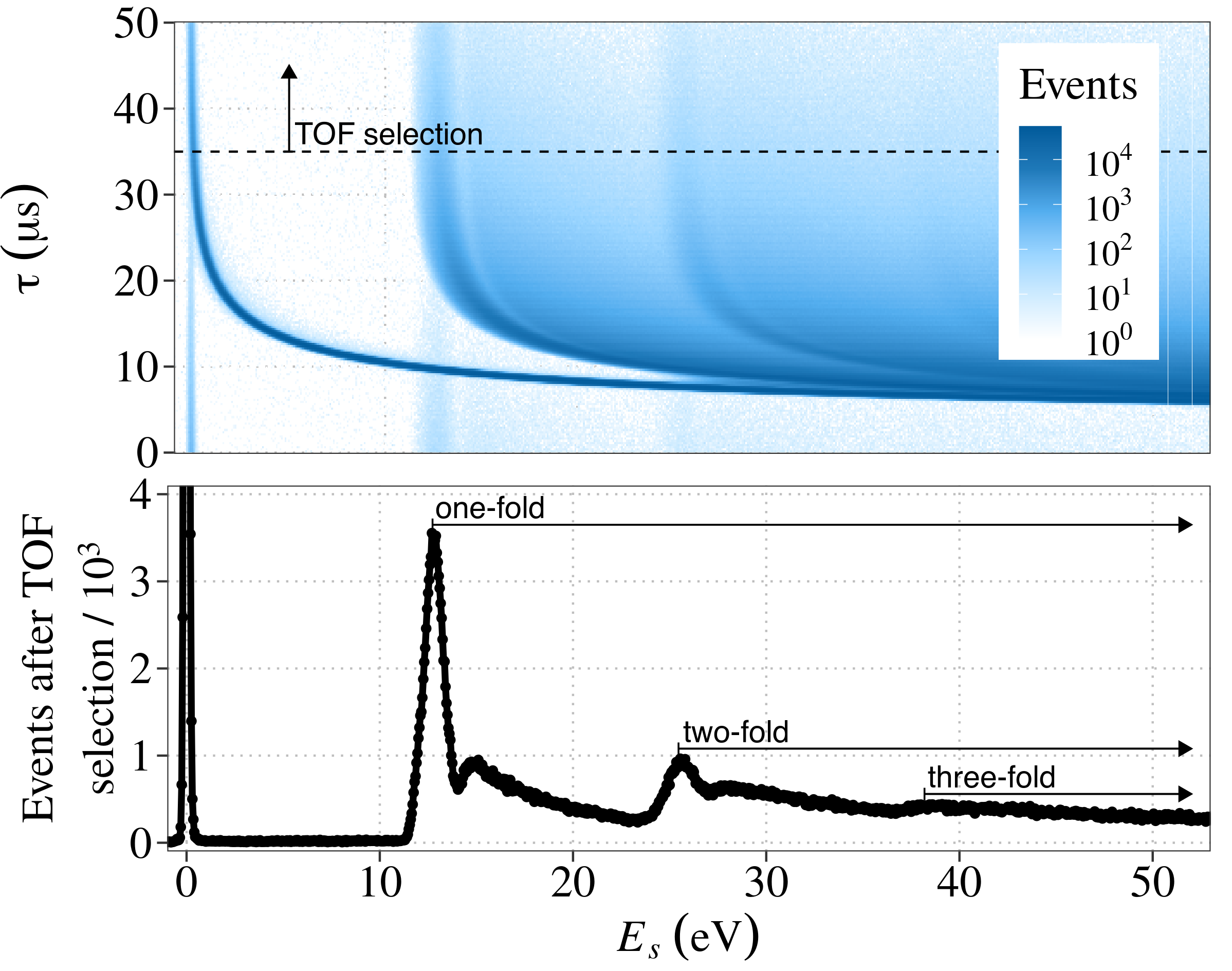}
	\caption{The differential measurements of the time of flight $\tau$ (top)
      and its one-dimensional projection on the electron surplus energy
      $E_\mathrm{s}$ axis (bottom) at \SI{86}{\percent} of nominal column
      density.  The dashed line marks the lower boundary of the TOF selection at $\tau_{\mathrm{cut}} =
      \SI{35}{\micro\second}$. The bottom panel shows all events in the
      TOF selection.}
	\label{fig:tof-eloss}
\end{figure}

For the differential measurements, the laser was pulsed at \SI{20}{\kilo\hertz}
to be able to distinguish flight times up to \SI{50}{\us} between the pulses
(see Fig.~\ref{fig:tof-eloss}). In this mode, an estimated 0.35 electron per
pulse are emitted. Measurements at four different column densities were
performed, which are listed in Tab.~\ref{tab:integralMeasurements}.
Figure~\ref{fig:tof-eloss} shows the measurements at \SI{86}{\percent} nominal
column density $\rho_0 d$ as an example. The top panel shows the time-of-flight
versus surplus energy. Here the unscattered electrons as well as one-fold and
two-fold scattered electrons are prominently visible as hyperbolic structures.

A TOF selection of events with flight times longer than $\tau_{\mathrm{cut}} =
\SI{35}{\micro\second}$ is applied to obtain a differential spectrum, which is
projected on $E_\mathrm{s}$ and shown in the bottom panel. $\tau_{\mathrm{cut}}$
is chosen such that an energy resolution of $\approx\,$\SI{0.02}{\electronvolt}
is achieved. Higher $\tau_{\mathrm{cut}}$ allows for a higher energy resolution
but results in significantly lower statistics. The vertical features --- at
\SIlist{0;12.5;25}{\electronvolt} --- for ${\tau < \SI{25}{\micro\s}}$ are
electrons with flight times $>\,$\SI{50}{\micro\second} from a previous laser
pulse. These events are neglected in the analysis.

All events with $\tau$ in the range of \SI{35}{\micro\second} to
\SI{50}{\micro\second} are selected and corrected for laser intensity
fluctuations analogous to the integral analysis. The energy scale for each
measurement is constructed using the measured ramping speed of the high voltage and
the position of the peak of unscattered electrons set to $E_\mathrm{s} = 0$.

%% file: pileup.tex
\subsection{Pile-up correction}
\label{sec:pileup}

The focal-plane detector is optimized to count single-e\-lec\-tron events with an
energy resolution of $\Delta
E_\mathrm{FPD}\approx\SI{2}{\kilo\electronvolt}$. Due to the high electron rate
of the electron gun ($\approx\,$\SI{e4}{cps}) and the use of a single detector
pixel, pile-up effects become relevant. Furthermore, the pulsed electron beam
with $<\,$\SI{18}{\ns} FWHM windows creates a non-Poisson time distribution compared
to a constant wave light source.

\begin{figure}[tbp]
	\centering
	\includegraphics[width=\figWidth\linewidth]{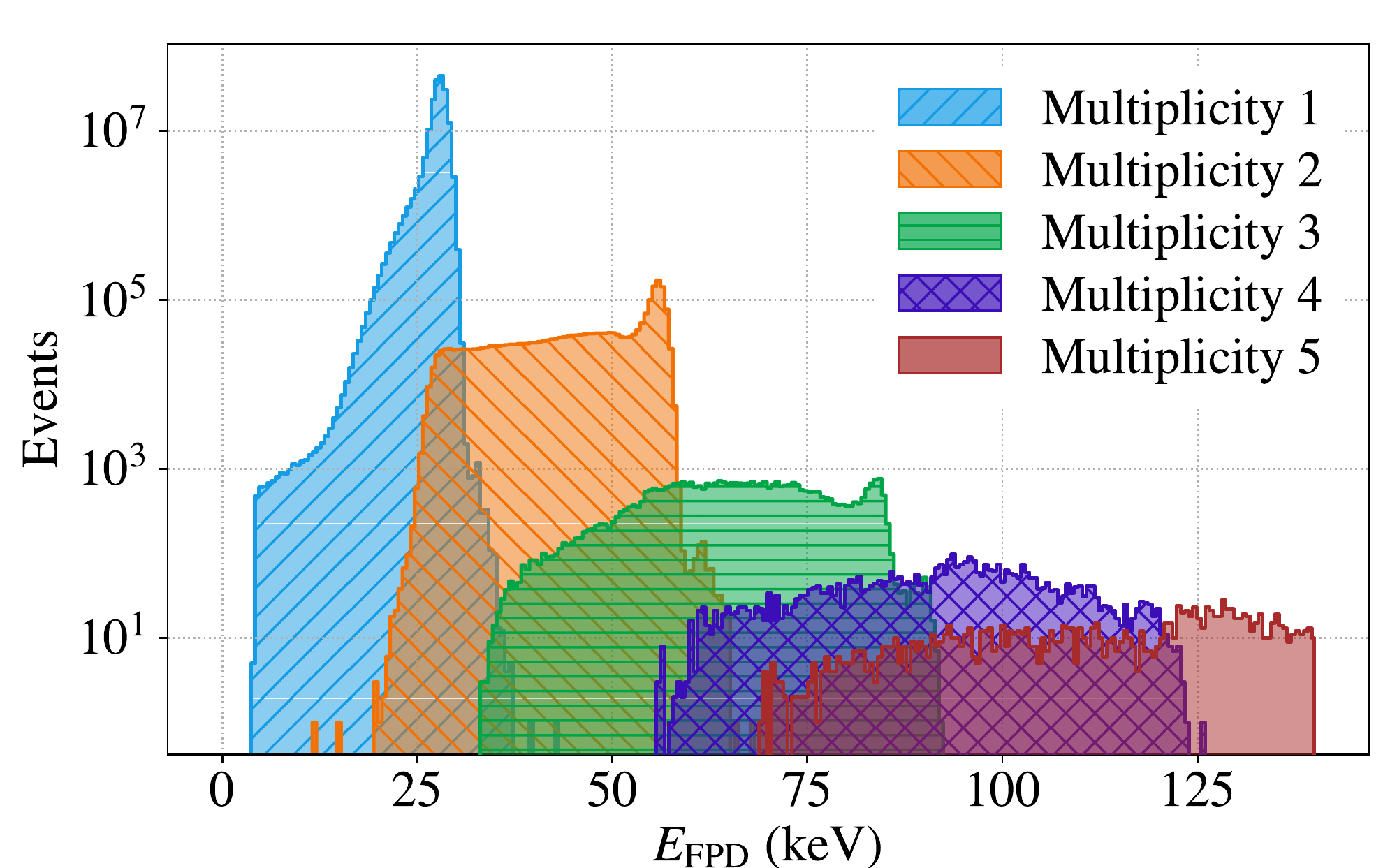}
	\caption{Reconstructed event energy in the focal-plane Si-detector for all events accumulated during the integral response function measurements at \SI{86}{\percent} nominal column density $\rho_0d$. The decomposition with the dedicated pile-up correction method shows that the different multiplicity regions overlap. This effect does not allow for a simple pile-up correction based on event energy alone. }
	\label{fig:energyHistogram}
\end{figure}

The electron flight time depends on the retarding potential and the energy loss
from scatterings inside the WGTS. The time difference of the electrons from the same pulse arriving at
the detector is thus modified as a function of the surplus energy. For
arrival-time differences shorter than the shaping time ($L=\SI{1.6}{\us}$) of the
trapezoidal filter used for pulse shaping of the detector signal, the electrons
are counted as one single event with correspondingly higher \linebreak event energy
$E_\mathrm{FPD}$ (Fig.~\ref{fig:energyHistogram}). The number of electrons within the same detector event is denoted as event multiplicity $\mathcal{M}$. As the peaks for
different multiplicity $\mathcal{M}$ events overlap in the
$E_\mathrm{FPD}$ histogram, a simple estimation of $\mathcal{M}$ based on
$E_\mathrm{FPD}$ is not possible. Processing the event signal with two
additional stages of trapezoidal filters allows more information on the signal
shape, such as the bipolar width $\mathcal{W}$ (i.e. the time difference of two consecutive zero crossings of the third trapezoidal-filter output), to be
obtained \cite{Aker2021design}. Electrons with arrival-time differences close to the shaping time distort the trapezoidal output of the first filter stage and thus change the determined bipolar width as a function of the arrival-time difference.
With the additional information on the pulse shape, these ambiguities
can be resolved and $\mathcal{M}$ can be estimated. The multiplicity estimate
$\bm\hat{\mathcal{M}}(E_\mathrm{FPD}, \mathcal{W})$ is obtained from Monte Carlo
simulations of the detector response for random combinations of $\mathcal{M}$
electrons arriving within the shaping time $L$. The estimate
$\bm\hat{\mathcal{M}}(E_\mathrm{FPD}, \mathcal{W})$ is not necessarily identical to
$\mathcal{M}$ as there are still remaining ambiguities, which are considered in
the uncertainty propagation (see Sec.~\ref{sec:MCpropagation}).  In the case of
the integral measurement data, the correction is made by weighting each event
with the estimator value. For the differential measurements, no pile-up
correction is required, but a $\bm\hat{\mathcal{M}}(E_\mathrm{FPD}, \mathcal{W})>1$ cut is
applied for background suppression (see Sec.~\ref{sec:background}).

\begin{figure*}[tbp]
	\centering
	\includegraphics[width=\figWidth\linewidth]{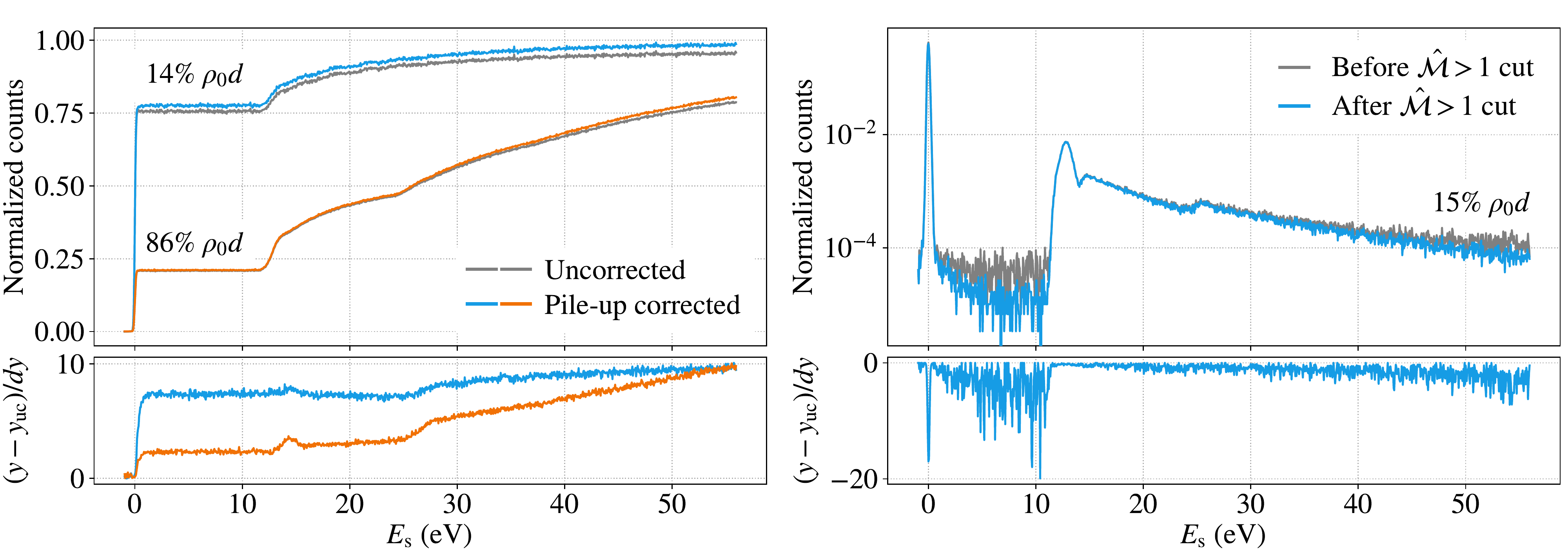}
	\caption{Left: Selection of measured response functions in integral mode
      before (grey line) and after (colored line) pile-up correction. The
      correction removes spectral shape distortions up to ten times larger than
      the statistical uncertainties.  Right: Differential response function
      before and after applying the $\bm\hat{\mathcal{M}}>1$
      cut. The cut reduces the background component by up to a factor of two
      without significantly influencing the shape of the signal component.
      Bottom: The difference between the uncorrected/uncut (uc) data and
      the corrected/cut data normalized to the data point uncertainties $dy$.
    }
	\label{fig:pileupcorrectedvsuncorrectedat50cd}
\end{figure*}

A comparison of the integral response function before and after pile-up
correction is provided in Fig.~\ref{fig:pileupcorrectedvsuncorrectedat50cd} to
demonstrate its dependence on the surplus energy at two different values of $\rho d$.
The dependence of $\bm\hat{\mathcal{M}}(E_\mathrm{FPD}, \mathcal{W})$ on the kinetic
energy of the electrons over the measurement range of \SI{60}{\eV} is neglected
in the correction and an average estimate is used instead.  The uncertainty due
to the correction method was evaluated with a full simulation of the detector
response for each of the response functions measured in integral mode. This
yields a correction stability at \num{5e-4}, which is considered as a systematic
uncertainty for the \linebreak energy-loss function determination.

%% file: background.tex
\subsection{Backgrounds}
\label{sec:background}
\begin{figure}[tbp]
	\centering
	\includegraphics[width=\figWidth\linewidth]{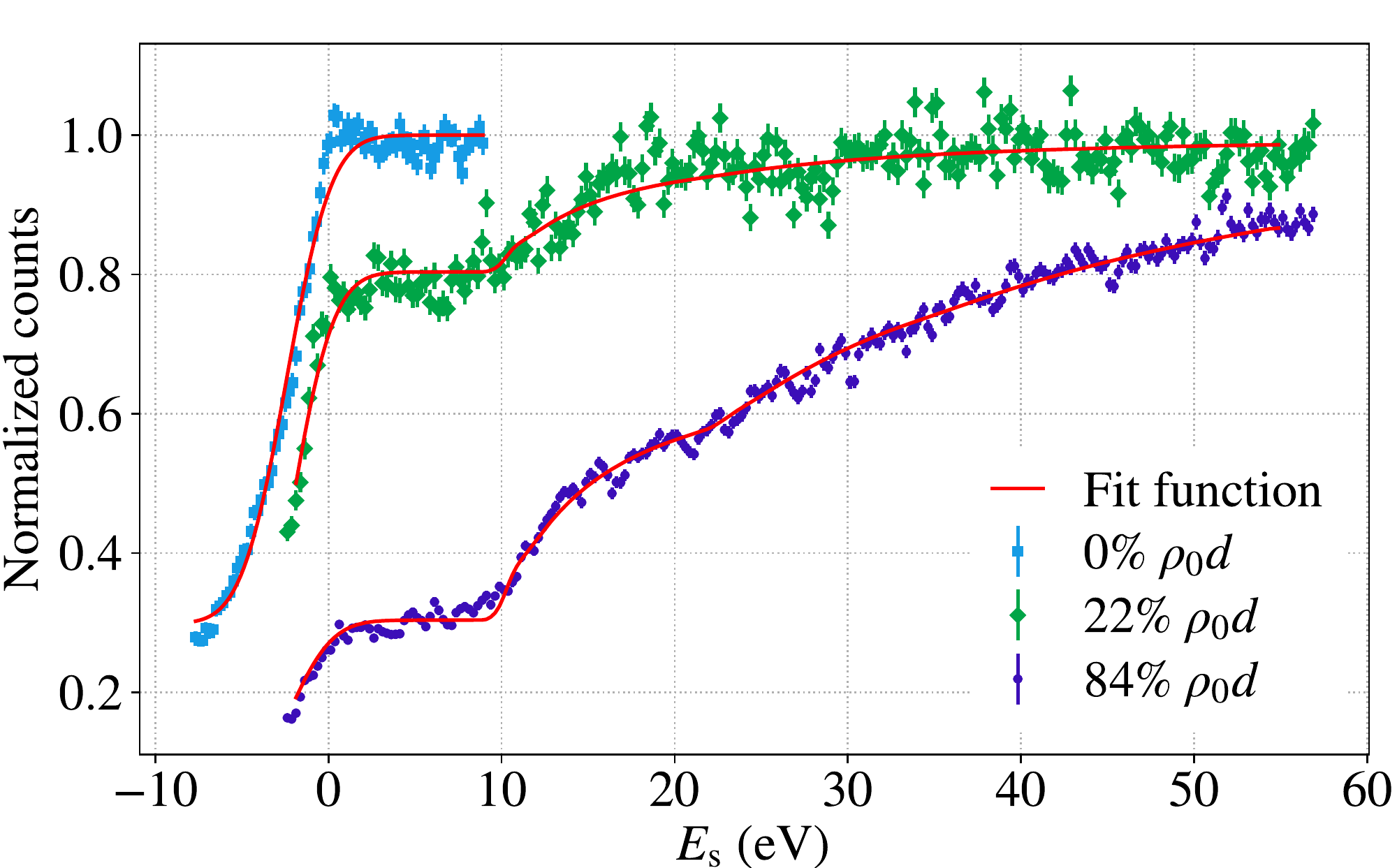}
	\caption{Background measurements of the electron gun with the light source
      turned off at different fractions of the nominal
      column density $\rho_0 d$.  Background electrons
      generated on the emission electrode of the electron gun show similar
      energy and column density dependencies as signal electrons.
      Compared to signal electrons, the background energy distribution
      is broader and shifted towards higher initial values. A combined fit to
      the data (red line) is used to determine the mean position
      $m_\mathrm{Bg}$ and the width $w_\mathrm{Bg}$ of the initial energy distribution. For better illustration, the
      shown data is normalized such that the region of unscattered electrons in
      the plateau at $E_\mathrm{s}\in\left[\SI{0}{\eV},\SI{8}{\eV}\right]$ equals
      $P_0(\mu)$.
		\label{fig:BackgroundMeasurements}}
\end{figure}

As the rear section is directly connected to the WGTS, tritium migration upstream
towards the electron gun cannot be completely prevented. Tritium can decay
within the acceleration fields of the electron gun. Ions created from the
$\upbeta$-decays are accelerated towards the photocathode, where their impact
can generate multiple secondary electrons simultaneously. Those electrons are
accelerated to the same energy as the signal photoelectrons. The kinetic energy
of the background electrons changes along with the change of the photocathode voltage $U_\mathrm{ph}$ in a scan. This results in a background spectrum following
the shape of an integral response function, as it is shown in
Fig.~\ref{fig:BackgroundMeasurements}. The background electrons only differ in
their initial energy distribution and the emission multiplicity (i.e. the number
of electrons generated from an ion impact). The mean energy $m_\mathrm{Bg}$ and
the Gaussian width $w_\mathrm{Bg}$ of the initial energy distribution of the secondary 
electrons can be obtained by performing a combined fit to the three background measurements 
using the same integral response-function model as described in \linebreak  Sec.~\ref{sec:Analysis}. 
The initial energy distribution dominates the spectral shape of the transmission function $T(E_\mathrm{s})$, which describes the transmission probability of the electrons inside the main spectrometer as a function of the surplus energy $E_\mathrm{s}$. The transmission function can be approximated with an error function using $m_\mathrm{Bg}$ and $w_\mathrm{Bg}$ as free parameters. The nine energy-loss function parameters were fixed to preliminary evaluated values during the fit. The best-fit result yields
\begin{equation}
\label{eq:backgroundParameters}
	m_\mathrm{Bg}=\SI{2.42\pm0.03}{\eV} \quad\text{and}\quad w_\mathrm{Bg}=\SI{ 2.05\pm0.04}{\eV}\,.
\end{equation}
The electron multiplicity distribution of the ion-induced events follows a
Poisson distribution (including ion-induced events with no electrons being
emitted) with the mean value
\begin{equation}
\label{eq:initialBackgroundMultiplicity}
	\bm\hat S=\SI{1.3\pm0.4}{}\,.
\end{equation}

\begin{figure}[tbp]
	\centering
	\includegraphics[width=\figWidth\linewidth]{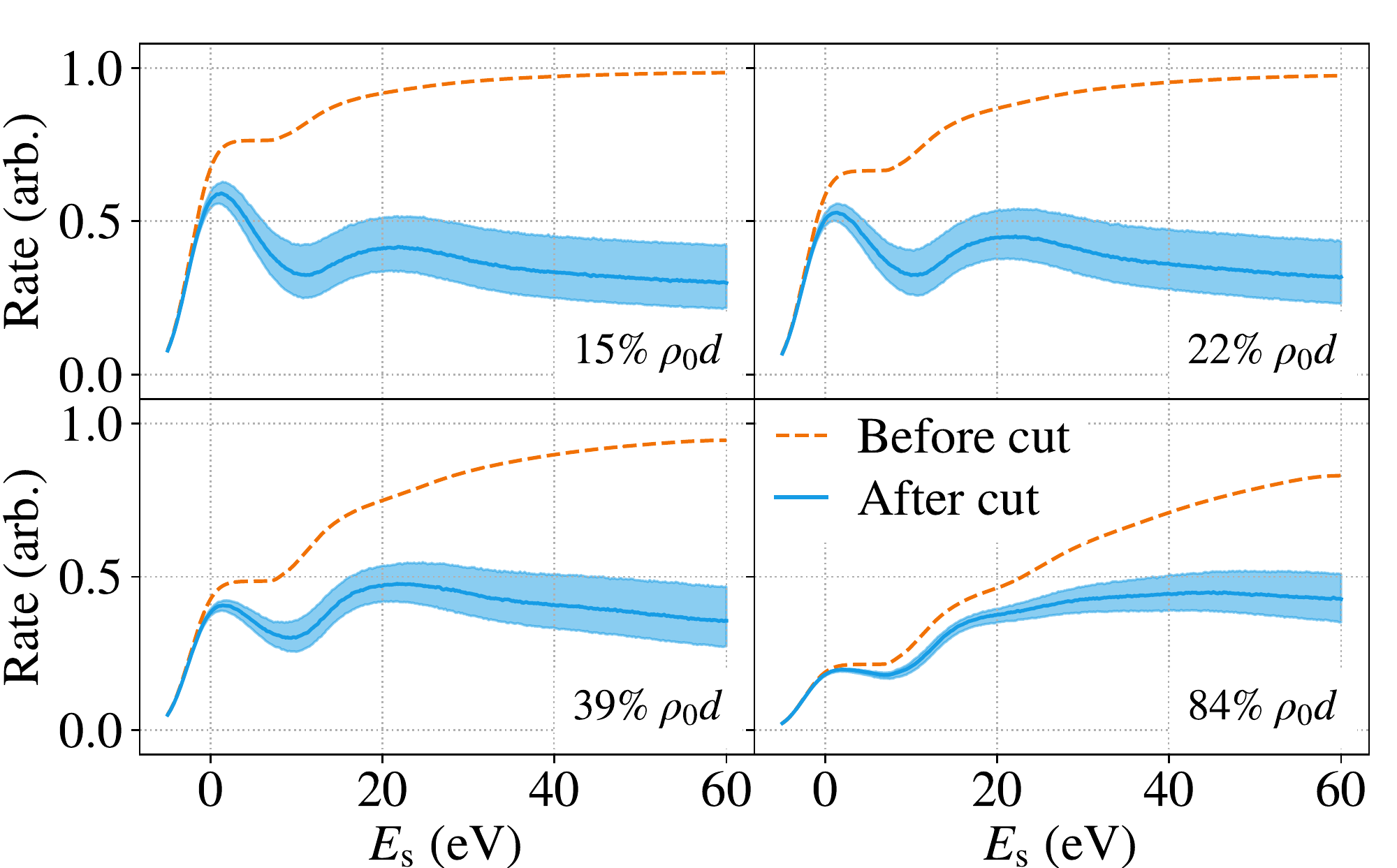}
	\caption{Simulated background spectra with and without multiplicity $\bm
      \hat \mathcal{M}>1$ cut for the differential mode after applying the TOF selection. The background spectra
      without multiplicity cut (orange) show the shape of an integral response
      function (cf. Fig.~\ref{fig:dataOnly_integral}). The TOF selection (not
      shown) does not affect this shape of background electrons. The background
      spectra after multiplicity cut (blue) are strongly reduced but are
      deformed in shape.  The shaded areas show the $1\sigma$ intervals
      resulting from the uncertainty on the mean emission multiplicity $\bm\hat S$ of the
      ion events provided in Eq.~\eqref{eq:initialBackgroundMultiplicity}.}
	\label{fig:backgroundAfterCut}
\end{figure}

Background events cause a larger detector pile-up effect compared to the signal
electrons generated by the pulsed laser, especially in the differential
data. The remaining events after the TOF selection are nearly unaffected by detector
pile-up, since only the scattered electrons survive. As the arrival time of
scattered electrons is delayed compared to other unscattered electrons from the
same light pulse, they do not arrive at the detector in time coincidence with
other electrons. This allows the multiplicity estimator
$\bm\hat{\mathcal{M}}(E_\mathrm{FPD}, \mathcal{W})$ to discriminate background events from
signal electrons. By excluding events with $\bm\hat{\mathcal{M}}>1$ in the analysis, the
background component can be reduced by about a factor of two without any
significant distortion of the signal component. A comparison of the differential response
function (at 15\% $\rho_0 d$) before and after applying the multiplicity cut is
provided in Fig.~\ref{fig:pileupcorrectedvsuncorrectedat50cd}, showing the
reduction of the background component. However, the multiplicity
$\bm\hat{\mathcal{M}}>1$ cut causes a distortion of the shape of the
background component, which is determined from simulations. The resulting
response functions of the background component after an event multiplicity
cut are displayed in Fig.~\ref{fig:backgroundAfterCut}.  The four
simulated spectra of the background components for the individual column
densities are included in the fit model.

%% file: fitResults.tex
\section{Analysis}
\label{sec:Analysis}
The energy-loss parameters in Eq.~\eqref{eq:katrinFitModel} are extracted with a
$\chi^2$-fit to multiple datasets in integral and differential mode at different column
densities. The systematic uncertainties in the energy-loss function (for example, those
due to the measurement conditions, pile-up and background effects) are determined
with Monte Carlo simulations (cf. Sec.~\ref{sec:MCpropagation}). Results are
given for molecular tritium and deuterium source gases below.

\subsection{Combined fit of the datasets}
\label{sec:combinedFitmodel}
The fit model is constructed with the energy-loss function, effects of multiple
scatterings in the source, energy smearing in the experimental setup, and the described background component.
The energy-loss function describes a single electron scattering. The probability
for $n$-fold scattering follows a Poisson distribution and is given by
\begin{equation}
P_n(\mu)=\frac{\mu^n}{n!}\exp\left(-\mu \right)\, ,
\end{equation}
with the expected mean number of scatterings $\mu$ given by
\begin{equation}
\mu=\rho d\cdot \sigma^{\mathrm{tot}}_{\mathrm{inel}}(qU_0).
\end{equation}
$\rho d$ is the column density during the individual measurements and $\sigma^\mathrm{tot}_\mathrm{inel}$ is the total inelastic
scattering cross section. To correct for the inelastic scattering cross section
at different kinetic energies, the parameter $\mu$ is scaled by the ratio
${\sigma^\mathrm{tot}_\mathrm{inel}(E_\mathrm{kin})/\sigma^\mathrm{tot}_\mathrm{inel}(qU_0)}$,
which gives $P_n(\mu, E_\mathrm{s})$. The effects of elastic scattering off
tritium can be neglected since the amount of energy transferred in these
scattering processes (${\overline{\Delta
  E}_{\mathrm{el}}=\SI{2.3}{\milli\electronvolt}}$ \cite{Kleesiek2019}) is
negligible compared to the energy smearing caused, among others, by the width of
the kinetic energy distribution of the electrons produced with the electron gun or the finite energy resolution of the KATRIN main spectrometer. The
experimental response to electrons that have been scattered $n$ times in the
source gas is given by the $n$-fold convolution of the energy-loss function
$f(\Delta E)$ with itself and convolved one time with the experimental transmission function
$T(E_\mathrm{s})$, leading to the following definition of the corresponding
scattering functions $\epsilon_n(E_\mathrm{s})$
\begin{align}
\label{eq:scattering_functions}
\epsilon_0(E_\mathrm{s}) =\; & T(E_\mathrm{s}) \; , \notag\\ 
\epsilon_1(E_\mathrm{s}) =\; & T(E_\mathrm{s}) \otimes f(\Delta E) \; , \notag\\
\epsilon_2(E_\mathrm{s}) =\; & T(E_\mathrm{s}) \otimes f(\Delta E) \otimes f(\Delta E) \; , \; ...\; ,
\end{align}
with $E_\mathrm{s}$ being the surplus energy of the electrons (see \linebreak Eq.~\eqref{eq:surplusEnergy}) and $\Delta E$ being the energy loss resulting from an inelastic
scattering.  The shape of the ionization tail of the energy-loss function is
corrected for the shape distortion \linebreak ($<\,$\SI{e-2}{\percent}) caused by the change
of the kinetic energy.

The model $R(E_\mathrm{s},\mu)$, which is fit to data, is the sum of the
scattering functions $\epsilon_n(E_\mathrm{s})$ weighted by the corresponding
Poissonian probabilities
\begin{equation}
\label{eq:responseFunction}
  R(E_\mathrm{s},\mu) = \sum_{n=0}^{4} P_n(\mu, E_\mathrm{s}) \cdot \epsilon_n(E_\mathrm{s}).
\end{equation}
Given that the surplus energies considered in the energy-loss analysis are
limited to ${E_\mathrm{s} \leq \SI{56}{\eV}}$, the highest scattering order that
needs to be considered is $n=4$.

\begin{figure}[tbp]
 \centering
 \includegraphics[width=\linewidth]{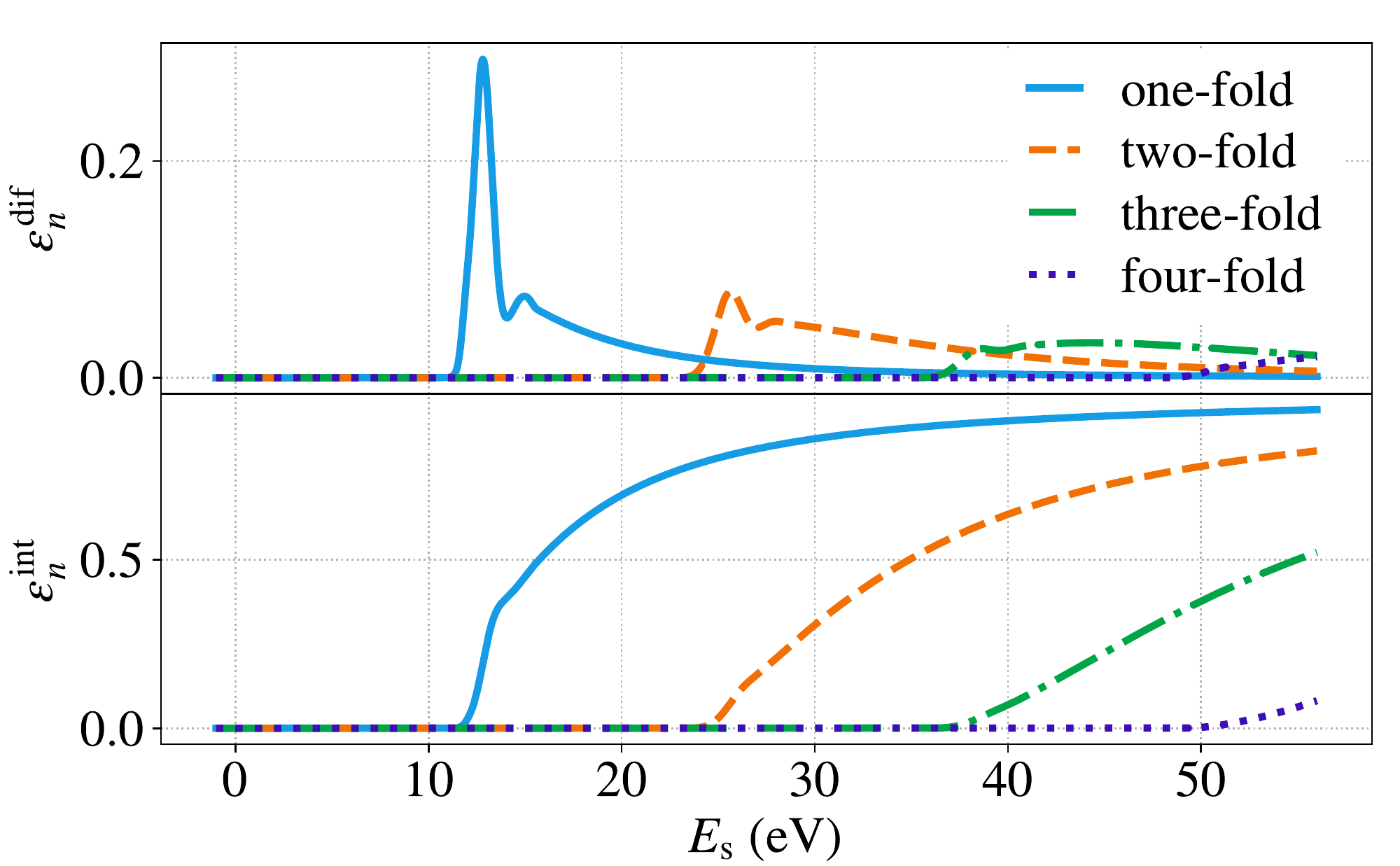}
 \caption{Differential ($\epsilon_{n}^{\mathrm{dif}} (E_\mathrm{s})$) and integral ($\epsilon_{n}^{\mathrm{int}} (E_\mathrm{s})$) scattering functions for up to four-fold scattering.
 } 
 \label{fig:scattering_functions}
\end{figure}

In the integral measurement, the shape of the experimental transmission function
$T_{\rm int}(E_\mathrm{s})$ is obtained from the response function with an empty
source volume; Eq.~\eqref{eq:responseFunction} collapses to $R(E_\mathrm{s}, 0)=
T(E_\mathrm{s})$. $T(E_\mathrm{s})$ is modeled with an error
function. Similarly, the transmission function for the differential data $T_{\rm
  dif}(E_\mathrm{s})$ could be obtained from a TOF measurement with an empty
source. However, it is simply given by the shape of the peak of unscattered
electrons observed at non-zero column densities; no additional measurement
is required in this case. Thus, we directly use the measurement data to
construct the fit model. Figure~\ref{fig:scattering_functions} shows the
scattering functions constructed for the differential ($\epsilon_{n}^{\rm
    dif} (E_\mathrm{s})$) and the integral ($\epsilon_{n}^{\rm int}
(E_\mathrm{s})$) measurement modes for the first four scattering orders.

In addition to the nine parameters in the energy-loss mo\-del in
Eq.~\eqref{eq:katrinFitModel} (amplitude, mean and width of the three Gaussians
contained in the model), several nuisance parameters are included in the
combined fit to differential and integral datasets taken at different column
densities. These nuisance parameters include normalization factors
$c^{\rm{dif(int)}}_{i}$, mean scattering probabilities $\mu^{{\rm dif(int)}}_{ i}$,
and background amplitudes \linebreak $b^{{\rm dif(int)}}_{i}$ for each differential (integral) dataset that is added
to the fit. In the fit, we minimize the following $\chi^2$ function for the
vector of free fit parameters $\vec{\mathcal{P}}$

\vskip 2cm
\begin{widetext}
\begin{align}
\label{eq:chisquare}
  \chi^2\left(\vec{\mathcal{P}}\right) = &  \sum_i^{N_{\rm dif}} \sum_j \left( \frac{c^{{\rm dif}}_{i} \, R^{\rm dif}(E_{\mathrm{s}, j},\mu^{{\rm dif}}_{ i}) + b^{{\rm dif}}_{i} \, B^{\rm dif}(E_{\mathrm{s}, j},\mu^{{\rm dif}}_{ i}) - y^{{\rm dif}}_{i,j}}{dy^{{\rm dif}}_{i,j}} \right)^2 \notag \\
   + & \sum_i^{N_{\rm int}} \sum_j \left( \frac{c^{{\rm int}}_{ i} \, R^{\rm int}(E_{\mathrm{s}, j},\mu^{{\rm int}}_{ i}) + b^{{\rm int}}_{ i}\, B^{\rm int}(E_{\mathrm{s}, j},\mu^{{\rm int}}_{ i}) - y^{{\rm int}}_{i,j}}{dy^{{\rm int}}_{i,j}} \right)^2  \notag \\
   + & \left(\frac{\int^{E_\mathrm{max}}_0 f(\Delta E) d(\Delta E) - 1}{\delta}\right)^2 \; ,
\end{align}
\end{widetext}

where $N_{\rm dif(int)}$ are the number of differential
(integral) datasets considered. $y^{{\rm dif(int)}}$ and $dy^{{\rm dif(int)}}$
represent the individual data points and their uncertainties. The index of summation $j$ denotes the data points of the individual datasets.  The first summand
of Eq.~\eqref{eq:chisquare} describes the contribution of the differential datasets to
the $\chi^2$ value.  The fit range for the differential datasets extends from
\SI{10}{\electronvolt} to \SI{56}{\electronvolt}, excluding the zero-scatter
peak and the adjacent background region, which do not contain information on the
energy-loss function. The second summand describes the contribution of integral
datasets with the fit range of \SI{-1}{\electronvolt} to \SI{56}{\electronvolt} \footnote{This extended fit range is required to determine the amplitude of the background component, which is only accessible below the transmission edge at $E_\mathrm{s}=\SI{0}{\eV}$.}. 
The ion-induced background component (see Sec.~\ref{sec:background}) is
considered in both summands. For the integral measurements, the shape of the
background component {$B^{\rm int}(E_{\mathrm{s}, j},\mu^{{\rm int}}_{ i})$} is
described by an integral response function (see
Fig.~\ref{fig:BackgroundMeasurements}), but with a different initial energy
distribution than the signal electrons. For the differential measurement, $B^{\rm
  dif}(E_{\mathrm{s}, j},\mu^{{\rm dif}}_{ i})$ is more complex and is obtained from
simulations described in Sec.~\ref{sec:background} and depicted in
Fig.~\ref{fig:backgroundAfterCut}.
The third summand is a pull term that ensures a proper normalization of the
fitted energy-loss function up to $E_\mathrm{max}=(E-E_\mathrm{i})/2$ with a
desired precision of $\delta = 10^{-4}$.

\begin{figure*}[tbp]
\centering
 \includegraphics[width=\figWidth\linewidth]{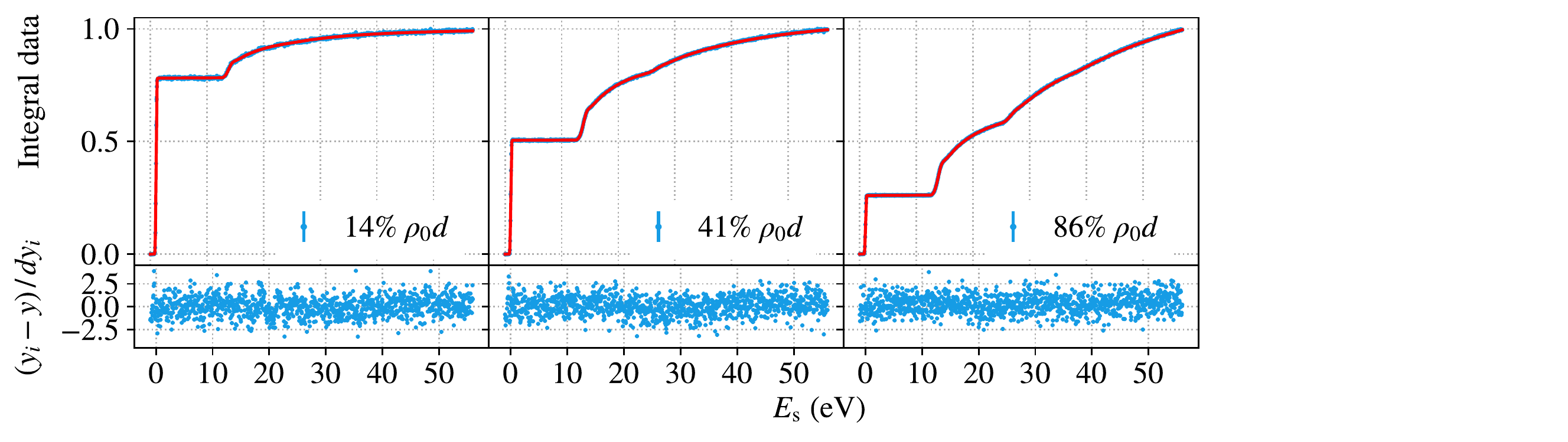}\newline  \includegraphics[width=\figWidth\linewidth]{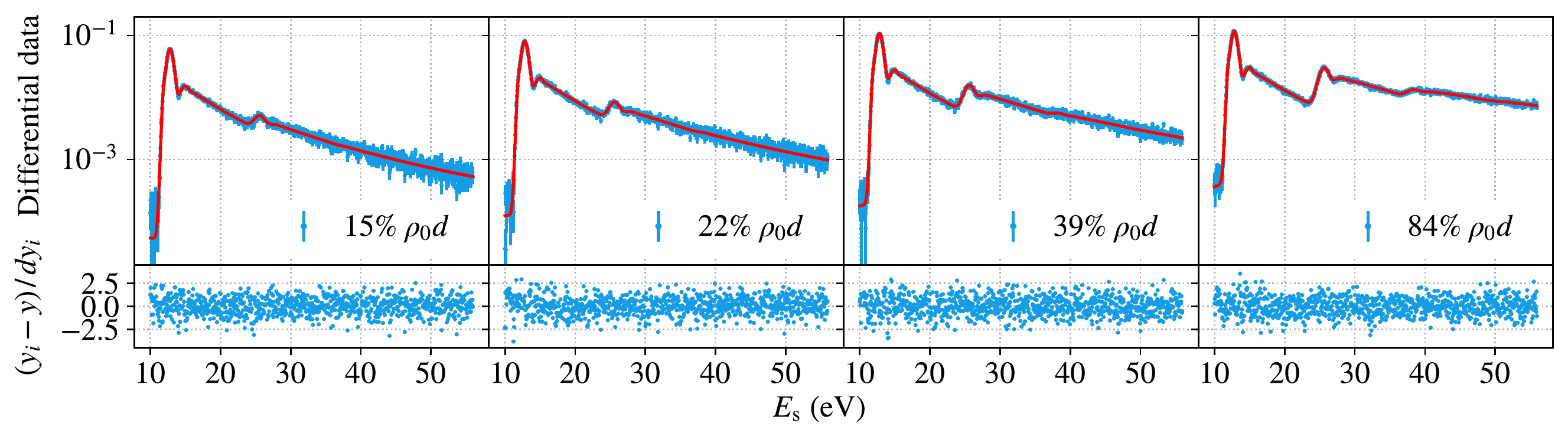}
 \caption{Results of the combined fit to the differential and integral datasets at different
   column densities. Each panel shows the data points (blue) and the best-fit
   result (red) in the upper part and the corresponding residuals in the lower
   part.  A normalization is applied to each of the differential and integral response
   functions. The differential data is normalized by the total number of counts within
   the fit range and the integral data by the number of counts in the last bin.
}
 \label{fig:combined_fit}
\end{figure*}

With the definition of the $\chi^2$ given in Eq.~\eqref{eq:chisquare}, a
combined fit to four differential datasets and three integral datasets taken at different
column densities (see Tab.~\ref{tab:integralMeasurements}) was performed. The
results are displayed in Fig.~\ref{fig:combined_fit} for each of the differential and
integral datasets included in the fit.

\begin{figure}[tbp]
	\centering
	\includegraphics[width=\figWidth\linewidth]{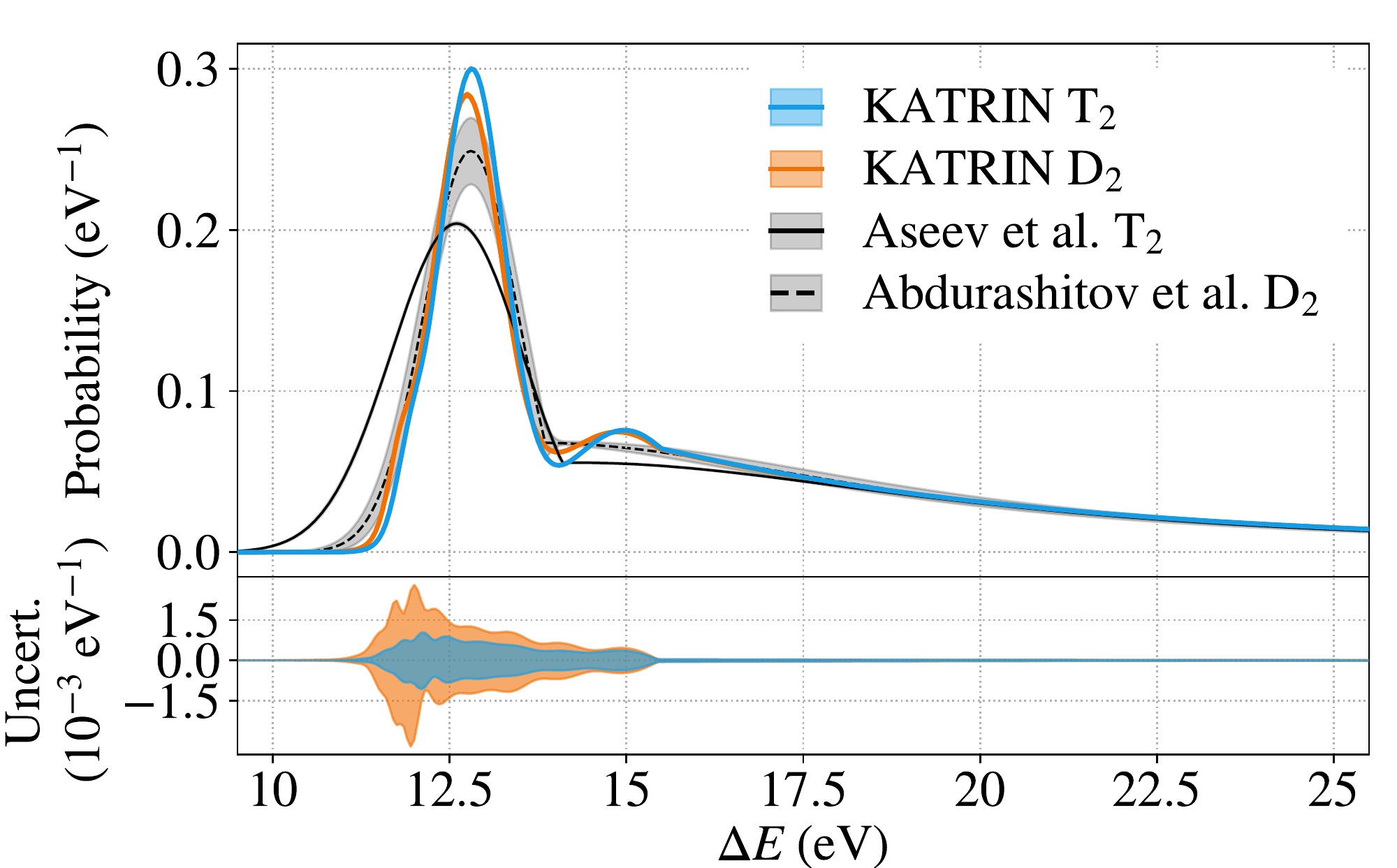}
	\caption{A comparison of the energy-loss functions in D$_2$ and T$_2$ from
      this work and previous measurements of Aseev et al. \cite{Ase00} and
      Abdurashitov et al. \cite{Abd17}. The y-axis indicates the
      probability density normalized in $\Delta E\in\left[0, E_\mathrm{max}\right]$ for energy losses $\Delta E$ due to inelastic
      scattering.  The Gaussian $1\sigma$ uncertainty bands are indicated by the shaded areas\protect\footnotemark[3]. Since the uncertainty of the KATRIN T$_2$ and D$_2$ results are too small to be visible in the top plot, the uncertainties are additionally shown in absolute values in the bottom plot.}
	\label{fig:elossFunctionComparison}
\end{figure}

The corresponding best-fit parameters of the energy-loss function are given in
Tab.~\ref{tab:fitResults}. The fit has a reduced $\chi^2$ value of
\SI{1.13\pm0.02}{}. A deviation from $\chi^2/N_\mathrm{dof}=1$ can arise from an imperfect
semi-empirical parametrization of the energy-loss function or an underestimation
of uncertainties. We do not observe significant structures in the fit residuals
in Fig.~\ref{fig:combined_fit} and thus inflate the uncertainties of the data
points by \linebreak $\sqrt{\chi^2/N_\mathrm{dof}}$ to achieve a $\chi^2/N_\mathrm{dof}=1$ \cite{PDG2020}.  The
statistical uncertainties from the fit are included in the third column of Tab.~\ref{tab:fitResults} with the
covariance matrix shown in Tab.~\ref{tab:covarianceMatrix} in the
Appendix.  Compared to the empirical energy-loss models of Aseev et al. and
Abdurashitov et al. superimposed on our results in
Fig.~\ref{fig:elossFunctionComparison}, the KATRIN result provides a better
energy resolution and reduced uncertainties.
As a consistency check, we extrapolate the energy-loss function (fitted up to \SI{56}{\eV}) to ${E_\mathrm{max} = \SI{9.280}{\keV}}$ yielding a mean energy loss of
$\overline{\Delta E}(\mathrm{T}_2)=30.79(1)_\mathrm{fit}$\,eV,
which agrees well with the value of \SI{29.9\pm 1}{\eV} reported by Aseev et al. \cite{Ase00}.

\begin{table}[tbp]
	\centering
	\caption{Best-fit parameters for the energy-loss function in molecular
      tritium as described in Eq.~\eqref{eq:katrinFitModel}. Parameter
      correlations are provided as a covariance matrix in
      Tab.~\ref{tab:covarianceMatrix} in the Appendix.}
    \begin{tabularx}{\tabWidth\linewidth}{llc}
		\toprule
		Parameter    & Unit & Value\\
		\midrule
	    $m_{1}$& \si{\electronvolt} &\tablenum{11.9189\pm0.0083}\\
	    $m_{2}$& \si{\electronvolt} &\tablenum{12.8046\pm0.0021}\\
        $m_{3}$& \si{\electronvolt} &\tablenum{14.9677\pm0.0041}\\
		$\sigma_{1}$& \si{\electronvolt} &\tablenum{0.1836\pm0.0070}\\
		$\sigma_{2}$& \si{\electronvolt} &\tablenum{0.4677\pm0.0022}\\
		$\sigma_{3}$& \si{\electronvolt} &\tablenum{0.907\pm0.013}\\
		$a_{1}$& \si{\per\electronvolt} &\tablenum{0.0328\pm0.0012}\\
	    $a_{2}$& \si{\per\electronvolt} &\tablenum{0.29570\pm0.00068}\\
	    $a_{3}$& \si{\per\electronvolt} &\tablenum{0.07575\pm0.00037}\\
		\bottomrule
	\end{tabularx}
	
	\label{tab:fitResults}
\end{table}

%\footnotetext[3]{The uncertainty band of the Aseev et al. \cite{Ase00} result is significantly smaller than the uncertainty band of the Abdurashitov et al. \cite{Abd17} result. However, the position of the Gaussian kernel was fixed to \SI{12.6}{\electronvolt} in the analysis of Aseev et al.}\addtocounter{footnote}{1}

%% file: errorPropagation.tex
\subsection{Systematic uncertainties}
\label{sec:MCpropagation}

Systematic uncertainties are not included in the combined fit; they are determined
separately by a Monte Carlo (MC) simulation framework. The framework generates
many MC samples, each composed of a detailed simulation of all integral and
differential datasets. The systematic effects under investigations can be folded
into these MC sets individually, or combined, with or without statistical
fluctuation of the count rates included. The underlying response function, on
which the MC generation is based, is taken from the best-fit values given in
Tab.~\ref{tab:fitResults}.

The considered systematic uncertainties cover known effects that arise from the
measurement conditions and effects specific to the integral or differential
analysis. All systematic effects are shown in Tab.~\ref{tab:systematics}. Their
implementation in MC generation is described in the following.

%footnote from Fig 11 placed here to be on same side in two column style
\footnotetext[3]{The uncertainty band of the Aseev et al. \cite{Ase00} result is significantly smaller than the uncertainty band of the Abdurashitov et al. \cite{Abd17} result. However, the position of the Gaussian kernel was fixed to \SI{12.6}{\electronvolt} in the analysis of Aseev et al.}\addtocounter{footnote}{1}

\begin{table*}[tbp]
	\centering
	\caption{A list of systematic uncertainties. The listed systematics are
      investigated by MC simulations yielding the contribution to the total
      parameter uncertainties and the parameter shifts, which are displayed in
      Fig.~\ref{fig:toyMCParamDistr}. The area of the uncertainty band of the
      energy-loss function caused by the individual systematic effects relative to that of all systematic effects is provided
      in the second to last column. The resulting parameter shifts due to each
      systematic effect are quantified in the last column as the area of the
      absolute deviations of the nominal function and the function given by the
      parameter means of the systematic variation. See text for more
      details.
		 }
\setlength\extrarowheight{5pt}
	\begin{tabularx}{\linewidth}{>{\raggedright\arraybackslash}X >{\raggedright\arraybackslash}X >{\hsize=3cm\raggedright\arraybackslash}Xrr}
		\toprule

		Systematic effect  & Source of input & Input values & {$\frac{\int|\sigma_\mathrm{sys}|}{\int |\sigma_\mathrm{all}|}$} & {$\frac{\int |f_0 - f_\mathrm{sys}|}{\int |\sigma_\mathrm{all}|}$}\\

		\midrule
		 Transmission-function model & error function fit to reference measurement & ${m_\mathrm{E}=\SI{-0.2\pm2.2}{\meV}}$\newline ${w_\mathrm{E}=\SI{90\pm1}{\meV}}$&\tablenum{0.94} &  \tablenum{0.260}\\
		 Column-density drift      & throughput sensor &  \SI{<0.2}{\percent\per h}\newline modeling according to sensor data  &\tablenum{0.015} & \tablenum{0.023}\\
		 Rate drift      & measurement data &  \SI{<0.15}{\percent\per h}  & \tablenum{0.002} &\tablenum{0.004} \\
		 Background          & bg measurement & ${m_\mathrm{Bg}=\SI{2.42\pm0.03}{\eV}}$
		 \newline
		 ${w_\mathrm{Bg}=\SI{2.05\pm0.04}{\eV}}$ &\tablenum{0.032} & \tablenum{0.008}\\
 		 Multiplicity cut            & bg measurement and simulation  &  	${\bm\hat S=\SI{1.3\pm0.4}{}}$ &\tablenum{0.153} & \tablenum{0.327}\\
		 Pile-up correction            & simulation  &  max. \SI{0.05}{\percent} & \tablenum{0.166} & \tablenum{0.271}\\
 		 Binning                       & HV sweep  &  bin width of \SI{0.05}{\eV} & \tablenum{0.0} & \tablenum{0.110} \\
 		 \midrule 
 		 All systematics& & & \tablenum{1.0} & \tablenum{0.395}\\
		\bottomrule
	\end{tabularx}

	\label{tab:systematics}
\end{table*}

\begin{itemize}
	\item \textit{Transmission-function model} In order to obtain an analytical
      description of the integral transmission function $T(E_\mathrm{s})$ for the construction of
      the integral response-func\-tion model, an error function is fit to a
      reference measurement with an empty WGTS. 
      The error function models the electron's surplus energy threshold needed for transmission in the main spectrometer \linebreak $m_\mathrm{E}=\SI{-0.2\pm2.9}{\milli\electronvolt}$ and the energy spread \linebreak $w_\mathrm{E}=\SI{90\pm2}{{\milli\electronvolt}}$ due to the angular and energy distribution of the electron gun and the energy resolution of the main spectrometer.
      To
      investigate the uncertainty of this analytical model, MC samples of
      the measurements at different column densities were generated with
      $m_\mathrm{E}$ and $w_\mathrm{E}$ drawn from a multivariate normal
      distribution according to the best-fit values above with the correlation between them taken into account. No uncertainty on the transmission-function
      model was considered for the differential data, since the peak of the unscattered
      electrons from the measurement data is directly used as the transmission
      function.
	\item \textit{Column-density drift} As the scattering
      probability $P_n$ depends on the column density, drifts in the column density
      during the measurements can cause a distortion of the response function.
      During the measurements at \SI{41}{\percent} of the nominal column
      density, drifts on the order of \SI{0.2}{\percent\per h} were visible. The
      reduced stability was caused by CO and tritiated methane freezing inside the injection
      capillaries. The CO and the methane were generated from radiochemical reactions with the stainless-steel surface during the \mbox{burn-in} period of the first tritium operation \cite{Aker2021}. The column
      density is constantly monitored with a throughput sensor, which allows the drift to be modeled precisely in the simulations. To do so, a linear function
      $\rho(t)$ is fit to the sensor data, yielding the slope of the drift and the corresponding parameter uncertainty. This linear function is used to model the rate drift due to the column density drift with the slope sampled according to its uncertainty.
	\item \textit{Rate drift} The electron-production rate of the electron gun can drift due to changes in
      the work function or a possible degradation of the photocathode (e.g. by
      ion impacts).  The number of unscattered electrons is analyzed for each
      run after correcting for drifts in the light intensity and the column
      density to monitor for intrinsic long-term rate drifts. Although the rate drift is very
      small at $\mathcal{O}(\SI{0.1}{\percent\per\hour})$, the resulting drift
      is used to modulate the response functions accordingly.
	\item \textit{Background} A background component created from
      secondary electrons by ion impact on the photocathode (cf.
      Sec.~\ref{sec:background}) adds to the response functions. In the MC
      simulations, a background component is added with the parameters of the initial energy distribution of the background electrons (see
      Eq.~\eqref{eq:backgroundParameters}) sampled according to their
      uncertainties.
	\item \textit{Multiplicity cut} The event multiplicity \linebreak
      ${\bm\hat{\mathcal{M}}(E_\mathrm{FPD}, w)>1}$ cut distorts
      the background shape in the differential measurements (see
      Fig.~\ref{fig:backgroundAfterCut}) depending on the initial electron multiplicity
      $\bm\hat S$ (see Eq.~\eqref{eq:initialBackgroundMultiplicity}) of the ion
      impact.  In the MC
      simulations, the value of $\bm\hat S$ is sampled according to its Gaussian
      uncertainty determined from the measurement and the resulting background component is added to the
      differential data.  Distortions on the signal component from the photoelectrons
      due to the multiplicity cut were investigated by dedicated detector
      simulations and added to the differential response functions.
	\item \textit{Pile-up correction} Detector pile-up is a dominant systematic effect for the integral measurements and is corrected with the pile-up reconstruction method described in Sec.~\ref{sec:pileup}.
	The efficiency $\zeta(E_{s})$ of this pile-up correction method is determined with detector simulations for each data point. The simulated response functions are multiplied by $\zeta(E_{s})$ to include the remaining distortions after applying the pile-up correction. The efficiency $\zeta(E_{s})$ is varied according to the Gaussian uncertainty determined in detector simulations.
	\item \textit{Binning} The response functions are measured by continuously ramping the emission energy of the electron gun. For the data analysis, the continuous data stream is binned into \num{50}-\si{\meV} bins. This binning effect is included in the MC simulations.
\end{itemize}

\begin{figure*}[tbp]
	\centering
	\includegraphics[width=0.73\figWidth\linewidth]{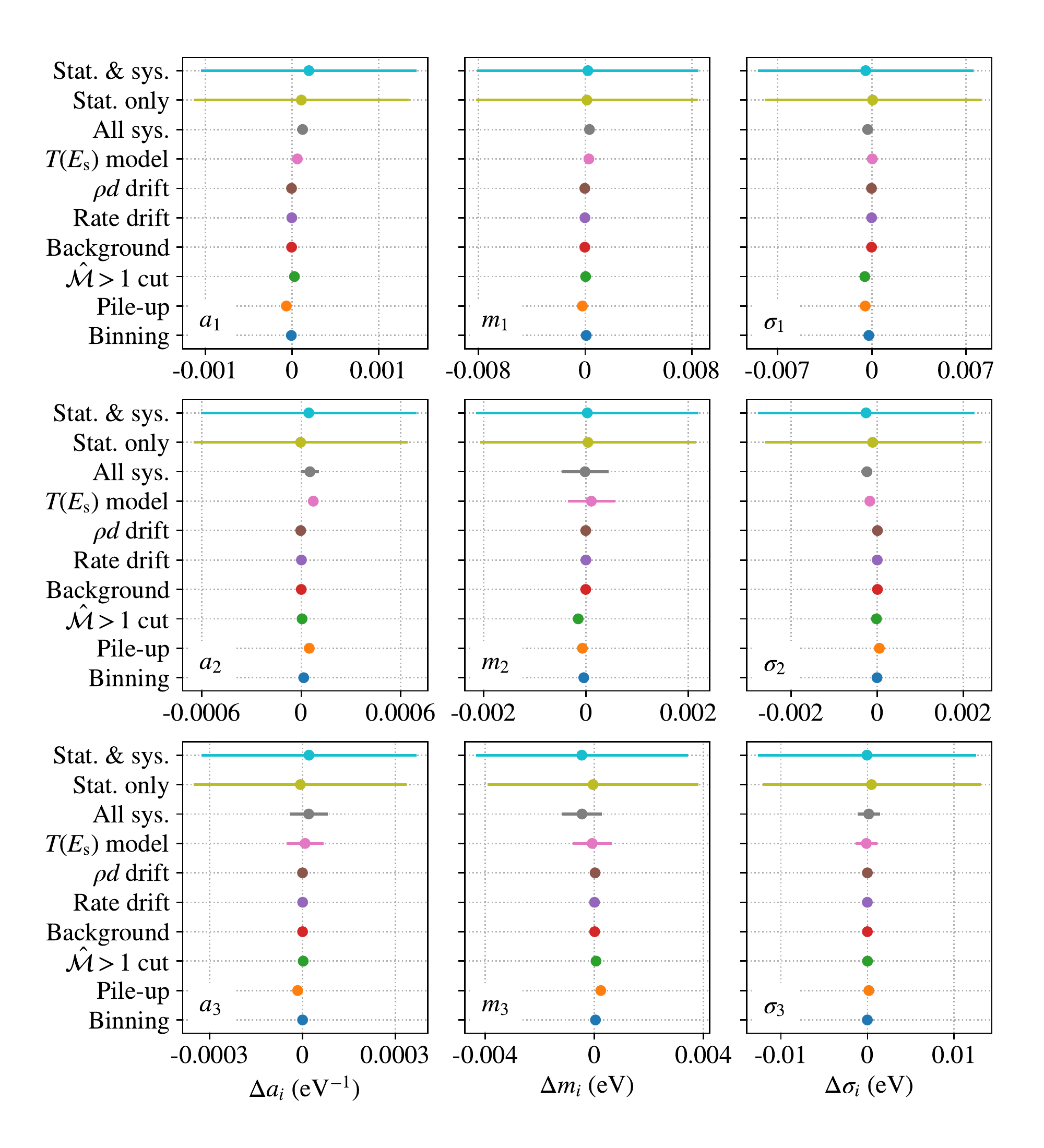}
	\caption{A breakdown of systematic uncertainties for all nine individual energy loss parameters as obtained from Monte Carlo simulations. Shown are the total uncertainty (stat. \& sys.), the statistical uncertainty (stat. only) as well as the total systematic uncertainty (all sys.). 
		The data points indicate the difference between the fit to data without any systematic effects and the median of the parameter distribution obtained from the fits to 10000 MC samples with systematic effects. The bars indicate the standard deviation of the distributions.
	The measurement is strongly dominated by the statistical uncertainty. The investigated systematic effects do not significantly contribute to a broadening of the parameter uncertainties nor to a significant shift of their mean values.
		\label{fig:toyMCParamDistr}}
\end{figure*}

A total of 10000 MC datasets are generated from the distributions of the systematic effects. Every MC dataset is fit and the best-fit values are taken to construct the probability distribution for each of the nine parameters of interest.
From these distributions, the parameter uncertainties are determined from the standard deviations. In addition, systematic parameter shifts are determined from the difference between the median of the distribution and the initial input value from the underlying energy-loss function.
The results of this evaluation are shown in Fig.~\ref{fig:toyMCParamDistr}. 
The total uncertainty is dominated by the statistics in the data and the widths of the distributions agree well with the parameter uncertainties of the best-fit result provided in Tab.~\ref{tab:fitResults}.
In order to condense the information of the nine parameter uncertainties for easier interpretation, two metrics are defined. They are shown in the last two columns of Tab.~\ref{tab:systematics}. 
The first metric, $\int|\sigma_\mathrm{sys}|/\int |\sigma_\mathrm{all}|$, is the area of the error band in the energy-loss function caused by the specific systematic ($\int|\sigma_\mathrm{sys}|$) with respect to the area of the error band caused by all systematic effects ($\int |\sigma_\mathrm{all}|$). The error bands originate from the combination of all nine parameter uncertainties. The areas of the error bands are estimates for the uncertainty of the scattering probability over the whole energy range.  
The second metric, 
$\int |f_0 - f_\mathrm{sys}|/\int |\sigma_\mathrm{all}|$, is the area of the difference between the nominal energy-loss function ($f_0$) and the energy-loss function ($f_\mathrm{sys}$) obtained from the simulations including the individual systematic uncertainties. This difference is normalized to $\int |\sigma_\mathrm{all}|$. A difference can be created by shifts of the nine parameter values caused by a given systematic effect. The impact of parameter shifts on the functional form of the energy loss is found to be smaller than the impact of the parameter uncertainties. The dominant contribution to the systematic uncertainty originates from the transmission-function model.
Since the total uncertainty of the energy-loss function is dominated by statistical uncertainties in the data and no significant parameter shifts are found, the considered systematic effects are negligible and not further considered in this study.
 
%% file: deuteriumResults.tex
\subsection{Deuterium results}
\label{sec:D2} 
Measurements, similar to the ones described in Sec.~\ref{sec:measurements}, were performed with molecular deuterium as source gas in an early commissioning run of the KATRIN experiment.
Four integral measurements at \SIlist{0;5;35;87}{\percent} of the nominal source density and a single differential measurement at \SI{5}{\percent} were made.
The data were processed and fit in the same manner as described in Secs.~\ref{sec:measurements} and \ref{sec:Analysis}.
For the combined $\chi^2$-fit of the deuterium measurements, the best-fit result is obtained at a reduced $\chi^2=\SI{1.57\pm0.02}{}$. Similar to the tritium data, the uncertainties of the data points are rescaled by $\sqrt{\chi^2/N_\mathrm{dof}}$ to obtain a reduced $\chi^2=1$. The parameter values as well as the covariance matrix are provided in Tabs.~\ref{tab:fitResultsD2} and \ref{tab:covarianceMatrixD2}.
The slightly increased $\chi^2$ value and the larger model uncertainties (cf. Fig.~\ref{fig:elossFunctionComparison}) can be explained by the presence of a stronger detector pile-up in the integral data due to an electron rate that was twice as high as that of the tritium measurements combined with the availability of only one differential dataset.
A full propagation of the systematic uncertainties was not performed for the deuterium measurements as the simulations for tritium showed that the measurements are strongly dominated by the statistical uncertainty. Furthermore, neither the systematic uncertainty due to methane freezing causing column-density drift nor the background generated from tritium ions is present in the absence of tritium.

Figure \ref{fig:elossFunctionComparison} shows the minor differences of the energy-loss models for deuterium and tritium, as the electronic excitation states are shifted to lower energies on the order of \SI{100}{\meV} \footnote{Such a difference between the different hydrogen isotopologs is theoretically expected. The observed difference of $\mathcal{O}(\SI{100}{\meV})$ is in agreement with preliminary calculations in dipole approximation in which the peak positions of the rovibrationally resolved spectra for the 2p$\sigma$ $^1\Sigma_\mathrm{u}$ and the 2p$\pi$ $^1\Pi_\mathrm{u}$ states were compared for the isotopes D$_2$ and T$_2$ \cite{Miniawy2021}.}. 
Extrapolating again to $E_\mathrm{max}$ the energy loss function results in a mean energy loss of\linebreak
$\overline{\Delta E}(\mathrm{D}_2)=30.64(1)_\mathrm{fit}$\,eV,
for the dominant deuterium isotopologs. This mean energy-loss value is \SI{0.15}{\eV} smaller than for tritium isotopologs, but we should not forget that we extrapolate the energy-loss function in energy by a factor 200 and we do not account for systematic uncertainties here for this consistency check\footnote{Just to get an order of magnitude estimate of the systematic uncertainties of the mean energy loss, we have left the junction point $E_\mathrm{i}$ between the three Gaussians and the BED tail in equation (\ref{eq:katrinFitModel}) free in our fits, yielding already an additional systematic uncertainty on the mean energy loss as big as the  discrepancy. We want to add that the systematics of this extrapolation is not critical for the determination of the energy-loss function in our interval of interest $[\SI{0}{\eV},\SI{54}{\eV}]$.}. 

\begin{table}[tbp]
	\centering
	\caption{
		Best-fit parameter values for the energy-loss function in molecular deuterium as described in Eq.~\eqref{eq:katrinFitModel}. Parameter correlations are provided as a covariance matrix in Tab.~\ref{tab:covarianceMatrixD2}.}
	\begin{tabularx}{\tabWidth\linewidth}{llc}
		\toprule
		Parameter    & Unit & {Value}\\
		\midrule
		$m_{1}$& eV &\tablenum{11.793\pm0.020}\\
		$m_{2}$& eV &\tablenum{12.7300\pm0.0046}\\
		$m_{3}$& eV &\tablenum{14.875\pm0.011}\\
        $\sigma_{1}$& eV &\tablenum{0.166\pm0.017}\\
		$\sigma_{2}$& eV &\tablenum{0.4828\pm0.0053}\\
		$\sigma_{3}$& eV &\tablenum{1.073\pm0.032}\\
		$a_{1}$& \si{\per\eV} &\tablenum{0.0344\pm0.0028}\\
		$a_{2}$& \si{\per\eV} &\tablenum{0.2737\pm0.0015}\\
		$a_{3}$& \si{\per\eV} &\tablenum{0.07466\pm0.00047}\\
		\bottomrule
	\end{tabularx}
	
	\label{tab:fitResultsD2}
\end{table}

%% file: summary.tex
\section{Summary and Outlook}
\label{sec:summary}
A series of precision measurements of the energy-loss function of \num{18.6}-\si{\keV} electrons scattering off molecular tritium and deuterium gas was performed.
The measurements were carried out in the KATRIN setup by using a pulsed beam of monoenergetic and angular selected electrons from a photoelectron source. The measurements were made in integral and differential time-of-flight measurement modes.

A new semi-empirical parametrization of the energy-loss function was developed, which describes the set of electronic states in combination with molecular excitations, dissociation, and ionization better than previous models.
This new model is described by nine parameters, which were determined by performing a combined $\chi^2$-fit to both integral and differential measurement data. The measurements and analyses performed in this work achieved a significant improvement over existing empirical energy-loss models in terms of energy resolution and uncertainties.
A detailed investigation of the systematic effects shows that the parameter uncertainties are dominated by statistical uncertainties. This allows further improvement in precision in future measurements. 

The obtained electron energy-loss function in tritium was used in the analysis of the first KATRIN dataset, which led to an improved upper limit of the effective neutrino mass ${m_\nu<\SI{1.1}{\eV}}$ (90\% C.L.) \cite{Aker2019}.
For this dataset, recorded at reduced source strength, the uncertainty of the energy-loss model contributes to the systematic uncertainty of the observable $m_\nu^2$ with $\sigma(m_\nu^2)<\SI{e-2}{\eV^2}$ and is inconsequential compared to other effects \cite{Aker2021}.
The achieved precision of the energy-loss function is close to the target effect of  ${\sigma(m_\nu^2)<\SI{7.5e-3}{\eV^2}}$ \cite{KAT04} that is necessary for reaching the final KATRIN sensitivity of ${m_\nu=\SI{0.2}{\eV}}$ (90\% CL).

%% file: appendix.tex
\FloatBarrier
\section*{Appendix}
\setlength\tabcolsep{2pt}
\setlength\extrarowheight{4pt}

\begin{table*}[!ht]
	\centering
	\caption{Covariance matrix for the parametrization of the energy-loss function for molecular tritium, as provided in Tab.~\ref{tab:fitResults}.}
	\resizebox{\linewidth}{!}{%
	\begin{tabular}{c|ccccccccc}
	\toprule
	   & $m_1$ &$m_2$ &$m_3$ &$\sigma_1$ &$\sigma_2$ &$\sigma_3$ &$a_1$ &$a_2$&$a_3$ \\ 
	\midrule
$m_1$&6.941$\cdot$10$^{\text{-5}}$&1.034$\cdot$10$^{\text{-5}}$&-3.388$\cdot$10$^{\text{-6}}$&4.537$\cdot$10$^{\text{-5}}$&-7.980$\cdot$10$^{\text{-6}}$&8.094$\cdot$10$^{\text{-6}}$&4.529$\cdot$10$^{\text{-6}}$&-6.505$\cdot$10$^{\text{-7}}$&-6.581$\cdot$10$^{\text{-8}}$\\
$m_2$&1.034$\cdot$10$^{\text{-5}}$&4.503$\cdot$10$^{\text{-6}}$&7.403$\cdot$10$^{\text{-7}}$&8.265$\cdot$10$^{\text{-6}}$&-1.206$\cdot$10$^{\text{-6}}$&-8.627$\cdot$10$^{\text{-6}}$&1.342$\cdot$10$^{\text{-6}}$&2.262$\cdot$10$^{\text{-7}}$&1.893$\cdot$10$^{\text{-7}}$\\
$m_3$&-3.388$\cdot$10$^{\text{-6}}$&7.403$\cdot$10$^{\text{-7}}$&1.641$\cdot$10$^{\text{-5}}$&-4.727$\cdot$10$^{\text{-6}}$&3.464$\cdot$10$^{\text{-6}}$&-2.255$\cdot$10$^{\text{-6}}$&-1.004$\cdot$10$^{\text{-6}}$&-1.272$\cdot$10$^{\text{-7}}$&-6.165$\cdot$10$^{\text{-7}}$\\
$\sigma_1$&4.537$\cdot$10$^{\text{-5}}$&8.265$\cdot$10$^{\text{-6}}$&-4.727$\cdot$10$^{\text{-6}}$&4.858$\cdot$10$^{\text{-5}}$&-8.929$\cdot$10$^{\text{-6}}$&1.503$\cdot$10$^{\text{-5}}$&2.481$\cdot$10$^{\text{-6}}$&-9.888$\cdot$10$^{\text{-9}}$&-1.840$\cdot$10$^{\text{-7}}$\\
$\sigma_2$&-7.980$\cdot$10$^{\text{-6}}$&-1.206$\cdot$10$^{\text{-6}}$&3.464$\cdot$10$^{\text{-6}}$&-8.929$\cdot$10$^{\text{-6}}$&4.746$\cdot$10$^{\text{-6}}$&-1.521$\cdot$10$^{\text{-5}}$&-1.755$\cdot$10$^{\text{-6}}$&-5.149$\cdot$10$^{\text{-7}}$&2.435$\cdot$10$^{\text{-7}}$\\
$\sigma_3$&8.094$\cdot$10$^{\text{-6}}$&-8.627$\cdot$10$^{\text{-6}}$&-2.255$\cdot$10$^{\text{-6}}$&1.503$\cdot$10$^{\text{-5}}$&-1.521$\cdot$10$^{\text{-5}}$&1.632$\cdot$10$^{\text{-4}}$&3.346$\cdot$10$^{\text{-6}}$&-2.017$\cdot$10$^{\text{-6}}$&-4.154$\cdot$10$^{\text{-6}}$\\
$a_1$&4.529$\cdot$10$^{\text{-6}}$&1.342$\cdot$10$^{\text{-6}}$&-1.004$\cdot$10$^{\text{-6}}$&2.481$\cdot$10$^{\text{-6}}$&-1.755$\cdot$10$^{\text{-6}}$&3.346$\cdot$10$^{\text{-6}}$&1.462$\cdot$10$^{\text{-6}}$&9.769$\cdot$10$^{\text{-8}}$&-4.513$\cdot$10$^{\text{-8}}$\\
$a_2$&-6.505$\cdot$10$^{\text{-7}}$&2.262$\cdot$10$^{\text{-7}}$&-1.272$\cdot$10$^{\text{-7}}$&-9.888$\cdot$10$^{\text{-9}}$&-5.149$\cdot$10$^{\text{-7}}$&-2.017$\cdot$10$^{\text{-6}}$&9.769$\cdot$10$^{\text{-8}}$&4.581$\cdot$10$^{\text{-7}}$&4.877$\cdot$10$^{\text{-8}}$\\
$a_3$&-6.581$\cdot$10$^{\text{-8}}$&1.893$\cdot$10$^{\text{-7}}$&-6.165$\cdot$10$^{\text{-7}}$&-1.840$\cdot$10$^{\text{-7}}$&2.435$\cdot$10$^{\text{-7}}$&-4.154$\cdot$10$^{\text{-6}}$&-4.513$\cdot$10$^{\text{-8}}$&4.877$\cdot$10$^{\text{-8}}$&1.354$\cdot$10$^{\text{-7}}$\\

\bottomrule
\end{tabular}
}
	\label{tab:covarianceMatrix}
\end{table*}
\begin{table*}[!ht]
	\centering
	\caption{Covariance matrix for the parametrization of the energy-loss function for molecular deuterium, as provided in Tab.~\ref{tab:fitResultsD2}.}
		\resizebox{\linewidth}{!}{%
	\begin{tabular}{c|*{9}{c}}
		\toprule
	   & $m_1$ &$m_2$ &$m_3$ &$\sigma_1$ &$\sigma_2$ &$\sigma_3$ &$a_1$ &$a_2$&$a_3$ \\ 
		\midrule
$m_1$&3.883$\cdot$10$^{\text{-4}}$&5.087$\cdot$10$^{\text{-5}}$&-2.607$\cdot$10$^{\text{-5}}$&2.487$\cdot$10$^{\text{-4}}$&-4.157$\cdot$10$^{\text{-5}}$&6.592$\cdot$10$^{\text{-5}}$&1.214$\cdot$10$^{\text{-5}}$&-4.525$\cdot$10$^{\text{-6}}$&-3.856$\cdot$10$^{\text{-7}}$\\
$m_2$&5.087$\cdot$10$^{\text{-5}}$&2.093$\cdot$10$^{\text{-5}}$&1.873$\cdot$10$^{\text{-5}}$&4.040$\cdot$10$^{\text{-5}}$&-2.989$\cdot$10$^{\text{-6}}$&-5.680$\cdot$10$^{\text{-5}}$&4.437$\cdot$10$^{\text{-6}}$&2.116$\cdot$10$^{\text{-6}}$&4.871$\cdot$10$^{\text{-7}}$\\
$m_3$&-2.607$\cdot$10$^{\text{-5}}$&1.873$\cdot$10$^{\text{-5}}$&1.144$\cdot$10$^{\text{-4}}$&-3.436$\cdot$10$^{\text{-5}}$&4.237$\cdot$10$^{\text{-5}}$&-2.466$\cdot$10$^{\text{-4}}$&-8.612$\cdot$10$^{\text{-6}}$&5.337$\cdot$10$^{\text{-6}}$&9.459$\cdot$10$^{\text{-7}}$\\
$\sigma_1$&2.487$\cdot$10$^{\text{-4}}$&4.040$\cdot$10$^{\text{-5}}$&-3.436$\cdot$10$^{\text{-5}}$&2.793$\cdot$10$^{\text{-4}}$&-4.330$\cdot$10$^{\text{-5}}$&6.404$\cdot$10$^{\text{-5}}$&-4.041$\cdot$10$^{\text{-6}}$&-7.999$\cdot$10$^{\text{-7}}$&-5.273$\cdot$10$^{\text{-8}}$\\
$\sigma_2$&-4.157$\cdot$10$^{\text{-5}}$&-2.989$\cdot$10$^{\text{-6}}$&4.237$\cdot$10$^{\text{-5}}$&-4.330$\cdot$10$^{\text{-5}}$&2.798$\cdot$10$^{\text{-5}}$&-1.050$\cdot$10$^{\text{-4}}$&-7.907$\cdot$10$^{\text{-6}}$&-2.660$\cdot$10$^{\text{-7}}$&6.033$\cdot$10$^{\text{-7}}$\\
$\sigma_3$&6.592$\cdot$10$^{\text{-5}}$&-5.680$\cdot$10$^{\text{-5}}$&-2.466$\cdot$10$^{\text{-4}}$&6.404$\cdot$10$^{\text{-5}}$&-1.050$\cdot$10$^{\text{-4}}$&1.033$\cdot$10$^{\text{-3}}$&1.829$\cdot$10$^{\text{-5}}$&-2.974$\cdot$10$^{\text{-5}}$&-1.231$\cdot$10$^{\text{-5}}$\\
$a_1$&1.214$\cdot$10$^{\text{-5}}$&4.437$\cdot$10$^{\text{-6}}$&-8.612$\cdot$10$^{\text{-6}}$&-4.041$\cdot$10$^{\text{-6}}$&-7.907$\cdot$10$^{\text{-6}}$&1.829$\cdot$10$^{\text{-5}}$&7.761$\cdot$10$^{\text{-6}}$&2.777$\cdot$10$^{\text{-7}}$&-6.118$\cdot$10$^{\text{-8}}$\\
$a_2$&-4.525$\cdot$10$^{\text{-6}}$&2.116$\cdot$10$^{\text{-6}}$&5.337$\cdot$10$^{\text{-6}}$&-7.999$\cdot$10$^{\text{-7}}$&-2.660$\cdot$10$^{\text{-7}}$&-2.974$\cdot$10$^{\text{-5}}$&2.777$\cdot$10$^{\text{-7}}$&2.173$\cdot$10$^{\text{-6}}$&4.225$\cdot$10$^{\text{-7}}$\\
$a_3$&-3.856$\cdot$10$^{\text{-7}}$&4.871$\cdot$10$^{\text{-7}}$&9.459$\cdot$10$^{\text{-7}}$&-5.273$\cdot$10$^{\text{-8}}$&6.033$\cdot$10$^{\text{-7}}$&-1.231$\cdot$10$^{\text{-5}}$&-6.118$\cdot$10$^{\text{-8}}$&4.225$\cdot$10$^{\text{-7}}$&2.193$\cdot$10$^{\text{-7}}$\\

		\bottomrule
	\end{tabular}
}
	\label{tab:covarianceMatrixD2}
\end{table*}
\FloatBarrier

%% file: ms.bbl
\begin{thebibliography}{10}
\providecommand{\url}[1]{{#1}}
\providecommand{\urlprefix}{URL }
\expandafter\ifx\csname urlstyle\endcsname\relax
  \providecommand{\doi}[1]{DOI \discretionary{}{}{}#1}\else
  \providecommand{\doi}{DOI \discretionary{}{}{}\begingroup
  \urlstyle{rm}\Url}\fi

\bibitem{Aker2021}
M.~Aker, et~al.
\newblock Analysis methods for the first {KATRIN} neutrino-mass measurement
  (2021)

\bibitem{KAT04}
{The KATRIN collaboration}, {KATRIN} design report.
\newblock {FZKA} scientific report 7090 (2005).
\newblock \doi{10.5445/IR/270060419}

\bibitem{Aker2021design}
M.~Aker, et~al.
\newblock The design, construction, and commissioning of the {KATRIN}
  experiment (2021).
\newblock \urlprefix\url{https://arxiv.org/abs/2103.04755}

\bibitem{Beamson1980}
G.~Beamson, H.~Porter, D.~Turner, J. Phys. E: Sci. Instrum. \textbf{13}(1), 64
  (1980).
\newblock \doi{10.1088/0022-3735/13/1/018}

\bibitem{LOBASHEV1985}
V.~Lobashev, P.~Spivak, Nucl. Instrum. Meth. A \textbf{240}(2), 305 (1985).
\newblock \doi{10.1016/0168-9002(85)90640-0}

\bibitem{Pic92}
A.~Picard, et~al., Nucl. Instrum. Meth. B \textbf{63}(3), 345 (1992).
\newblock \doi{10.1016/0168-583x(92)95119-c}

\bibitem{Friedel2019}
F.~Friedel, et~al., Vacuum \textbf{159}, 161 (2019).
\newblock \doi{10.1016/j.vacuum.2018.10.002}

\bibitem{Marsteller2021}
A.~Marsteller, B.~Bornschein, et~al., Vacuum \textbf{184}, 109979 (2021).
\newblock \doi{10.1016/j.vacuum.2020.109979}

\bibitem{Roettele2017}
C.~R\"{o}ttele, J. Phys. Conf. Ser. \textbf{888}, 012228 (2017).
\newblock \doi{10.1088/1742-6596/888/1/012228}

\bibitem{Ams15}
J.~Amsbaugh, et~al., Nucl. Instrum. Meth. A \textbf{778}, 40 (2015).
\newblock \doi{10.1016/j.nima.2014.12.116}

\bibitem{Kleesiek2019}
M.~Kleesiek, et~al., Eur. Phys. J. C \textbf{79}(3) (2019).
\newblock \doi{10.1140/epjc/s10052-019-6686-7}

\bibitem{Ase00}
V.N. Aseev, et~al., Eur. Phys. J. D \textbf{10}(1), 39 (2000).
\newblock \doi{10.1007/s100530050525}

\bibitem{Abd17}
D.N. Abdurashitov, et~al., Phys. Part. Nuclei Lett. \textbf{14}(6), 892 (2017).
\newblock \doi{10.1134/s1547477117060024}

\bibitem{Gei64}
J.~Geiger, Z. Physik \textbf{181}(4), 413 (1964).
\newblock \doi{10.1007/bf01380873}

\bibitem{Uls72}
R.C. Ulsh, et~al., J. Chem. Phys. \textbf{60}(1), 103 (1974).
\newblock \doi{10.1063/1.1680755}

\bibitem{Beh16}
J.~Behrens, et~al., Eur. Phys. J. C \textbf{77}(6) (2017).
\newblock \doi{10.1140/epjc/s10052-017-4972-9}

\bibitem{Kim2000}
Y.K. Kim, et~al., Phys. Rev. A \textbf{62}(5) (2000).
\newblock \doi{10.1103/physreva.62.052710}

\bibitem{Kim94}
Y.K. Kim, M.E. Rudd, Phys. Rev. A \textbf{50}(5), 3954 (1994).
\newblock \doi{10.1103/PhysRevA.50.3954}

\bibitem{Wec99}
P.~Weck, et~al., Phys. Rev. A \textbf{60}(4), 3013 (1999).
\newblock \doi{10.1103/PhysRevA.60.3013}

\bibitem{Pei_2002}
Z.~Pei, C.N. Berglund, Jpn. J. Appl. Phys. \textbf{41}(Part 2, No. 1A/B), L52
  (2002).
\newblock \doi{10.1143/jjap.41.l52}

\bibitem{PhDSack2020}
R.~Sack, {Measurement of the energy loss of 18.6\, keV electrons on deuterium
  gas and determination of the tritium Q-value at the {KATRIN} experiment}.
\newblock Ph.D. thesis, Westf\"alische Wilhelms-Universit\"at M\"unster (2020).
\newblock \urlprefix\url{http://nbn-resolving.de/urn:nbn:de:hbz:6-59069498754}

\bibitem{Steinbrink2013}
N.~Steinbrink, et~al., New J. Phys. \textbf{15}(11), 113020 (2013).
\newblock \doi{10.1088/1367-2630/15/11/113020}

\bibitem{Bonn1999}
J.~Bonn, et~al., Nucl. Instrum. Meth. A \textbf{421}(1), 256  (1999).
\newblock \doi{10.1016/S0168-9002(98)01263-7}

\bibitem{PDG2020}
P.~Zyla, et~al., Prog. Theor. Exp. Phys. \textbf{2020}(8) (2020).
\newblock \doi{10.1093/ptep/ptaa104}

\bibitem{Miniawy2021}
A.~{El~Miniawy}, A.~Saenz.
\newblock {\it private communication; to be published in the Bachelor thesis of
  A. El Miniawy} (2021)

\bibitem{Aker2019}
M.~Aker, et~al., Phys. Rev. Lett. \textbf{123}(22) (2019).
\newblock \doi{10.1103/physrevlett.123.221802}

\end{thebibliography}
